\documentclass[manuscript, nonacm]{acmart}

\usepackage{svg}

\AtBeginDocument{%
  \providecommand\BibTeX{{%
    \normalfont B\kern-0.5em{\scshape i\kern-0.25em b}\kern-0.8em\TeX}}}

\setcopyright{acmlicensed}
\copyrightyear{2024}
\acmYear{2024}
\acmDOI{XXXXXXX.XXXXXXX}

\acmConference[Conference acronym 'XX]{Conference}{Date}{Location}
\acmBooktitle{Conference,
 MonthName Month--Day, Year, City, State} 
\acmISBN{ISBN}

\usepackage{enumitem}
\newlist{researchquestions}{enumerate}{1}
\setlist[researchquestions]{label*=\textbf{RQ\arabic*}}
\begin{document}

\title{Rethinking Recommender Systems: Cluster-based Algorithm Selection}


\author{Andreas Lizenberger}
\affiliation{%
  \institution{University Siegen}
  \city{Siegen}
  \country{Germany}}
\email{andreas.lizenberger@student.uni-siegen.de}

\author{Ferdinand Pfeifer}
\affiliation{%
 \institution{University Siegen}
 \city{Siegen}
 \country{Germany}}
\email{ferdinand.pfeifer@student.uni-siegen.de}

\author{Bastian Polewka}
\affiliation{%
  \institution{University Siegen}
  \city{Siegen}
  \country{Germany}
\email{bastian.polewka@student.uni-siegen.de}
}

\renewcommand{\shortauthors}{Lizenberger, Pfeifer and Polewka}

\begin{abstract}
\textbf{Abstract.}
Cluster-based algorithm selection deals with selecting recommendation algorithms on clusters of users to obtain performance gains.
No studies have been attempted for many combinations of clustering approaches and recommendation algorithms.
We want to show that clustering users prior to algorithm selection increases the performance of recommendation algorithms.
Our study covers eight datasets, four clustering approaches, and eight recommendation algorithms. We select the best performing recommendation algorithm for each cluster.
Our work shows that cluster-based algorithm selection is an effective technique for optimizing recommendation algorithm performance. For five out of eight datasets, we report an increase in \textit{nDCG@10} between 19.28\% (0.032) and 360.38\% (0.191) compared to algorithm selection without prior clustering.
\end{abstract}




\maketitle

\section{Introduction}
In the digital age online platforms offer a vast number of products (items), such as movies \cite{netflix_report}, clothing \cite{zalando_report}, music \cite{spotify_report}, and many more. Online platforms want to encourage users to interact with as many items as possible, but online platforms can only display a fraction of items at any given time. Therefore, displaying items that match the preferences and tastes of users is important and motivates the integration of recommender systems. Recommender systems recommend a fraction of items based on previous interactions between users and items \cite{9216015}.
For example, Netflix deploys a recommender system to encourage users to watch more movies from their vast library of up to 18,214 movies and series \cite{netflix_report}. A user who frequently watches action movies is probably interested in the action genre. The recommender system deployed by Netflix recommends movies from the action genre to display the recommended movies to the user.

The performance of recommendation algorithms varies strongly which motivates algorithm selection for recommender systems \cite{beel2016towards}. Recommendation algorithms are the core of recommender systems and generate recommendations for users \cite{8550639}. The algorithm selection problem defines criteria for comparing a set of algorithms in order to select the best algorithm \cite{RICE197665}.
For example, applying different recommendation algorithms to news websites can result in a precision increase of up to 330\% (0.43) \cite{beel2016towards}.

The problem is selecting the single best recommendation algorithm for all users does not account for clusters of users. Clusters contain users with similar behavior and preferences. Clustering approaches detect these similarities and generate clusters of users \cite{xu2015comprehensive}. By selecting recommendation algorithms for each cluster we can leverage the unique strengths of each recommendation algorithm. We argue that cluster-based algorithm selection accounts for clusters of users resulting in an increase of overall recommendation performance.

Therefore, we want to determine whether selecting recommendation algorithms for clusters of users improves the overall recommendation performance of recommender systems. In our work, we aim to answer the following research question:
\begin{researchquestions}
\item What is the influence of clustering on the performance of recommendation algorithms?
\end{researchquestions}

To address our research question, we combine clustering with an already existing pipeline for training recommendation algorithms. For a total of eight datasets, we conduct a study on the recommendation performance achieved by selecting the best algorithms on each automatically generated cluster. \textbf{Our contribution} is cluster-based algorithm selection which increases the average \textit{nDCG@10} by 66.47\% compared to algorithm selection. We find combinations of clustering approaches and recommendation algorithms that significantly increase recommendation performance, answering \textbf{RQ1}.
Our long-term goal is to establish cluster-based algorithm selection.

Our implementation is publicly available on our GitHub repository\footnote{\url{https://code.isg.beel.org/PG_SS_23}} and
contains documentation for the reproducibility of our experiments.

\section{Related Work}
Cluster-based algorithm selection is still open to research. To the best of our knowledge, existing work covers a single clustering approach prior to algorithm selection. We review related work to cluster-based algorithm selection in the fields: clustering, meta-learning, automated recommender systems (AutoRecSys) and ensembling.

Alexander Nechaev et al.\ propose an approach to select algorithms based on user groups that is similar to our approach. The authors generate user groups by clustering with \textit{HDBSCAN}. \textit{HDBSCAN} clusters users with the following meta-features: relative count, standard deviation and skewness of user's ratings. The authors employ the following recommendation algorithms: \textit{SVD}, \textit{SVD++}, \textit{KNN Baseline}, \textit{Baseline Only}, \textit{Co-Clustering}. The authors select the best performing algorithm for each cluster \cite{meltsov2019utilizing}.

The influence of cluster-based algorithm selection on the performance of recommender systems is not researched extensively yet. The authors suggest a single clustering approach in combination with collaborative filtering. Cluster-based collaborative filtering attempts to improve adaptability and scalability of recommender systems while maintaining coverage and performance of recommendations. The problem is a marginal recommendation performance increase compared to collaborative filtering without prior clustering \cite{10.1145/2365952.2365997} \cite{GUO201514} \cite{10.1145/1076034.1076056} \cite{1241167}.

Meta-learning is an approach to deal with the algorithm selection problem. The idea of meta-learning is to analyze meta-information derived from datasets and performances of algorithms. The authors perform meta-learning algorithm selection on different levels: the global-level which covers datasets \cite{romero2013meta,10.1145/2611040.2611054,10.1007/978-3-319-46227-1_25, 10.1145/3240323.3240378}, the middle-level which covers groups of users or single users \cite{10.1145/2365952.2366002}, and the micro-level which covers user-item pairs \cite{edenhofer2019augmenting,collins2018novel}. 

AutoRecSys defines a pipeline to standardize and automate the stages of recommender systems to deal with the combined algorithm selection and hyperparameter optimization problem (CASH). AutoRecSys approaches deal with the CASH problem by training multiple algorithms with varying hyperparameters according to a search strategy, e.g. \textit{Random Search}, \textit{Bayesian optimization}, and selecting the best algorithm after hyperparameter optimization \cite{10.1145/3604915.3610656,10.1145/3383313.3411529,10.1145/3604915.3608886}. 

Ensembling is a technique to improve the performance of a recommender system. Jahrer et al.\ demonstrate that multiple recommendation algorithm perform better than a single recommendation algorithm \cite{10.1145/1835804.1835893}. Ensembling works with a broader range of user interactions when both explicit and implicit feedback is available \cite{6984809, 10.1145/2664551.2664556}. Ensembling is not only applicable to make recommendations to the users, but is also applicable in the context of detecting similarities of users \cite{forouzandeh2021presentation}. 

\section{Cluster-based AutoRecSys} \label{clusterBasedAutoRecSys}
\textit{LensKit-Auto} \cite{10.1145/3604915.3610656} is an AutoRecSys toolkit to automate the recommender systems pipeline. \textit{LensKit-Auto} offers preprocessing, hyperparameter optimization and algorithm training for the development of recommender systems. Hyperparameter optimization can be performed with a search strategy, e.g. \textit{Random Search} or \textit{Bayesian Optimization}. For each recommendation algorithm we use \textit{LensKit-Auto} to perform \textit{Random Search} to find optimized hyperparameters. We recommend based on the optimized hyperparameters. Optimizing hyperparameters enables our comparison to focus on the performance of the different recommendation algorithms rather than the choice of hyperparameters. We assess the performance of each recommendation algorithm by comparing either \textit{nDCG} or \textit{Precision}.

\begin{figure}[ht]
\centering
\includegraphics[width=1.0\columnwidth]{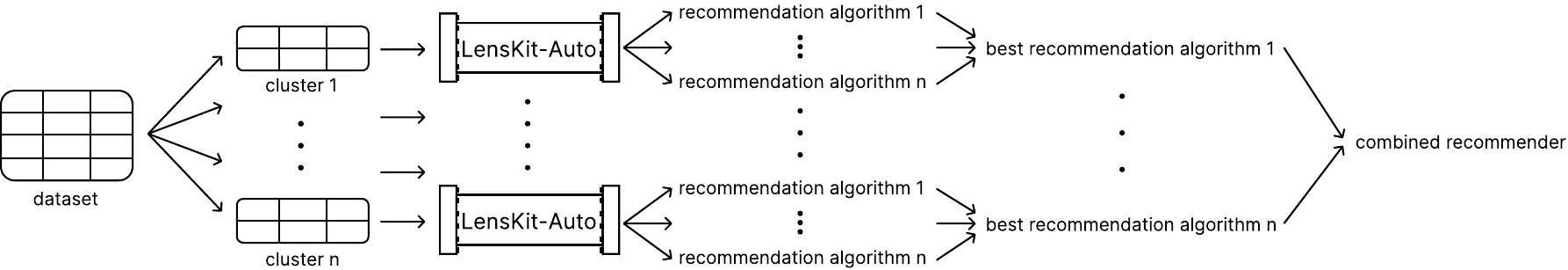}
\caption{Cluster-based AutoRecSys process.}
\label{fig:automatedClusteringProcess}
\end{figure}

We employ cluster-based AutoRecSys to find the best algorithms on each cluster to solve the cluster-based algorithm selection problem. First, we iterate over a list of clustering approaches and parameters to generate a set of disjoint clusters. Then, we apply the AutoRecSys library \textit{LensKit-Auto} to train recommendation algorithms on each cluster. For each cluster we select the best performing recommendation algorithm. We combine the selected best performing algorithms to a combined recommender (see Figure \ref{fig:automatedClusteringProcess}). Our approach can be classified as middle-level algorithm selection, because our approach operates on sets of clusters of users. We introduce our selected clustering approaches in the sections \ref{clusteringKMeans} and \ref{clusteringGraph}.


\subsection{Clustering with \texorpdfstring{$k$}{}-Means} \label{clusteringKMeans}
\(K\)-Means \cite{kmean} is a centroid-based clustering approach that generates clusters around centroids. Each centroid represents the center of a cluster. \(K\)-Means select \(k\) many centroids in order to generate \(k\) many clusters. For each user the distance to all centroids is computed. Users are assigned to a cluster based on the smallest computed distance to the centroid of the cluster. For each cluster the centroid is recomputed as the mean of all users of the cluster. The assignment and recomputing step are repeated until the cluster assignment from the current iteration is equal to the cluster assignment from the previous iteration \cite{kmean}.
For example, we want to cluster users by age. Given two clusters, the first cluster has a mean age of 20 and the second cluster a mean age of 40. A 25 year old user has a distance of 5 years to the centroid of the first cluster and a distance of 15 years to the centroid of the second cluster. Therefore, the 25 year old user is assigned to the first cluster with the mean age of 20. Once all users are assigned to a cluster, the mean age of the clusters is recomputed.

\paragraph{Clustering by Number of Interactions}
The first meta-information we choose for clustering with \(k\)-Means is the number of interactions between users and items. Number of interactions is the sum of all interactions that users have with items. We choose number of interactions for two reasons. The first reason is that generating clusters that separate users with a different number of interactions improves the recommendation score for users with many interactions \cite{beelPruning}. The second reason is that Jianling Wang et al.\ indicate that removing interactions from users with many interactions improves the recommendations for users with fewer interactions \cite{10.1145/3485447.3512147}.
For example, we have user A and B who both watched 40 movies and user C who watched 200 movies. Both user B and C share 20 of their watched movies with user A. We want to recommend a movie to user A. The recommendations provided by utilizing user B are more relevant to user A because user A and B share 50\% of their watched movies and user A and C share only 10\% of their watched movies.

\paragraph{Clustering by Item-Interaction Vector}
The second meta-information we choose for clustering with \(k\)-Means is the item-interaction vector of each user. The item-interaction vector encodes binarily whether users interacted with items or not and number of interactions as the last entry. For example, if a user interacted with item A, item B but not with item C, the item-interaction vector is \((1, 1, 0, 2)^T\).
Only considering number of interactions can lead to clusters with an insufficient number of users. The problem is that recommendation algorithms trained on clusters with an insufficient number of users perform poorly in our tests.
For example, clusters with an insufficient number of users can occur if we have a dataset with 10.000 users of which three users have 1000 interactions and remaining users have less than 20 interactions. Clustering based on number of interactions would result in a cluster with only three users, because the users have a similar number of interactions.
Clustering with item-interaction vector benefits \(k\)-Means, because additional meta-information is available to find similarities between users.
We demonstrate the availability of additional meta-information using an example of four users A, B, C, D, and two items A and B.
User A and B have an interaction with item A but not with item B, so the item-interaction vector is \((1, 0, 1)^T\).
User C and D have an interaction with item B but not with item A, so the item-interaction vector is \((0, 1, 1)^T\).
If we cluster the users based on number of interactions then the four users form a single cluster, because all users have same number of interactions.
With the item-interaction vector we can additionally cluster the users based on the items that the users have interacted with, resulting in a cluster with user A and B, and a cluster with user C and D.

\subsection{Clustering with Graphs} \label{clusteringGraph}
Our final two clustering approaches are the graph-based approaches \textit{Louvain Method for Community Detection} (\textit{Louvain}) \cite{louvain} and \textit{Greedy Modularity} \cite{greedy_modularity}. Graph-based clustering approaches analyze graphs and generate clusters by detecting densely interconnected subgraphs called communities. User-item interactions represent a graph: users and items are nodes, interactions are edges. 
\textit{Louvain} and \textit{Greedy Modularity} find communities based on modularity. Modularity measures the density of edges within communities compared to edges between communities. Initially, all nodes represent a community. Iteratively, \textit{Louvain} and \textit{Greedy Modularity} check for each node whether modularity is improved by moving the node to a neighboring community. The process of moving nodes to neighboring communities stops when the modularity cannot be further improved \cite{louvain, greedy_modularity}.

The meta-information we choose for graph-based clustering is a graph encoding of the user-item interaction matrix. We generate the graph by deriving all unique users and items. For every user and items we add a node to the graph. We add edges between users and items by traversing each user interaction entry and connecting the user and item contained in the entries. In order to obtain clusters using graph based approaches, we find communities by applying the graph-based approaches. The resulting communities consist of user and item nodes. We generate a cluster for each community by extracting all users from the community. The items are not relevant for our user-based clustering approaches, therefore we can ignore the item nodes.

\section{Methodology}
Our work analyzes the influence of clustering on algorithm selection. Each clustering approach generates a set of clusters. For each cluster in the generated set of clusters we evaluate the performance of recommendation algorithms. We select the best performing algorithm for each cluster. For each set of clusters we obtain a combined recommender that uses the best performing recommendation algorithm of each cluster. In order to measure the influence of clustering, we compare the combined recommender to a baseline. The baseline performance is the performance of the recommendation algorithm that performs best on the dataset prior to clustering. The performance of our combined recommender is the weighted sum of the best performing algorithm of each cluster. The algorithms are weighted by the number of users in the respective cluster divided by the number of all users in the dataset. We are interested in finding a combined recommender that outperforms the baseline.

The experiment was executed on a computer cluster, where each node has in total 64 cores from two AMD EPYC 7452 CPUs and 256 GB DDR4 3200MHz RAM.

\subsection{Design Decisions}
We explain the reasoning behind our parameter choices for our experimental pipeline. The clustering approaches utilize two clustering parameters: cluster count and resolution. Cluster count sets the number of clusters for the \textit{\(k\)-Means clustering approach}, while \textit{Greedy Modularity} uses it as an upper limit to generate clusters. In our experiments cluster count ranges from two to eight clusters. We reason the upper limit of eight clusters based on previous tests of our pipeline on an upper limit of 15 clusters. We measured no increase in performance with more than seven clusters. A higher cluster count could be feasible if the size of the dataset is larger than our tested datasets. \textit{Louvain} and \textit{Greedy Modularity} generate clusters based on resolution. Resolutions greater than one lead to many small clusters and resolutions smaller than one lead to few big clusters \cite{louvain, greedy_modularity}. In our experiments we choose resolutions 0.8, 0.9, default of 1.0, 1.1, and 1.2. With \textit{LensKit-Auto} we choose the \textit{Random Search} strategy for hyperparameter optimization, because \textit{Random Search} allows us to optimize for our two metrics \textit{nDCG} and \textit{Precision} simultaneously. We argue that with a limit of 100 iterations and a search limit of four hours, \textit{Random Search} will find enough optimized hyperparameters to compare the clustering approaches.

\subsection{Algorithms and Datasets} 
We use eight recommendation algorithms of which two are baseline algorithms and eight different datasets.
The recommendation algorithms are from the libraries \textit{LensKit} \cite{10.1145/3340531.3412778} and \textit{Implicit} \cite{githubImplicit}.
From \textit{LensKit} we use the algorithms: \textit{Random}, \textit{PopScore}, \textit{Item-Item Nearest Neighbour}, \textit{User-User Nearest Neighbour}, and \textit{Implicit Matrix Factorization}.
The algorithms from Implicit are: \textit{Alternating Least Square}, \textit{Bayesian Personalized Ranking}, and \textit{Logistic Matrix Factorization}.
The datasets come from a variety of domains, such as: Articles, Locations, Movies, Music and Socials. Table \ref{tab:datasets} shows all datasets with information about the number of users, items, interactions and domain of the data.
The implicit feedback datasets are: \textit{Globo} \cite{10.1145/3270323.3270328, 8908688}, \textit{Hetrec-Lastfm} \cite{10.1145/2043932.2044016}, \textit{Nowplaying} \cite{nowplaying-RS}, \textit{Retailrocket} \cite{Retailrocket} and \textit{Sketchfab} \cite{Sketchfab}. The explicit feedback data sets are: \textit{MovieLens-1M} \cite{10.1145/2827872}, \textit{MovieLens-100k} \cite{10.1145/2827872}, and \textit{MovieTweetings} \cite{MovieTweetings}.
All of the mentioned datasets are five-core pruned, so that every user and item has at least five interactions \cite{10.1145/3357384.3357895, 10.1145/3460231.3474275, 10.1145/3523227.3546770}.
Five-core pruning is necessary to prevent cold start problems with our recommendation algorithms.
We convert the explicit datasets into implicit datasets by dropping the rating column.

\begin{table}
\centering
\caption{Five-core pruned dataset statistics. Implicit datasets upper part, explicit datasets lower part.}
\label{tab:datasets}
\begin{tabular}{l|l|l|l|l|l|l|l}
Name             & \#Interactions & \#Users   & \#Items & Avg.\#Int./User & Avg.\#Int/Item & Sparsity & Domain    \\ \hline
Globo            & 2,482,163      & 157,926   & 11,832  & 15.72           & 209.78         & 99.87\%  & Articles  \\
Hetrec-Lastfm    & 71,355         & 1,859     & 2,823   & 38.38           & 25.28          & 98.64\%  & Music     \\
Nowplaying       & 2,447,318      & 64,392    & 95,277  & 38.01           & 25.69          & 99.96\%  & Music     \\
Retailrocket     & 240,938        & 22,178    & 17,803  & 10.86           & 13.53          & 99.94\%  & Shopping  \\
Sketchfab        & 547,477        & 25,655    & 15,274  & 21.34           & 35.84          & 99.86\%  & Social    \\ \hline
MovieLens-100k   & 81,697         & 943       & 1,203   & 86.64           & 67.91          & 92.8\%   & Movies    \\ 
MovieLens-1M     & 835,789        & 6,038     & 3,307   & 138.42          & 252.73         & 95.81\%  & Movies    \\
MovieTweetings   & 563,309        & 20,643    & 8,810   & 27.29           & 63.94          & 99.69\%  & Movies    \\
\end{tabular}
\end{table}

\begin{figure}[ht]
\centering
\includegraphics[width=0.6\columnwidth]{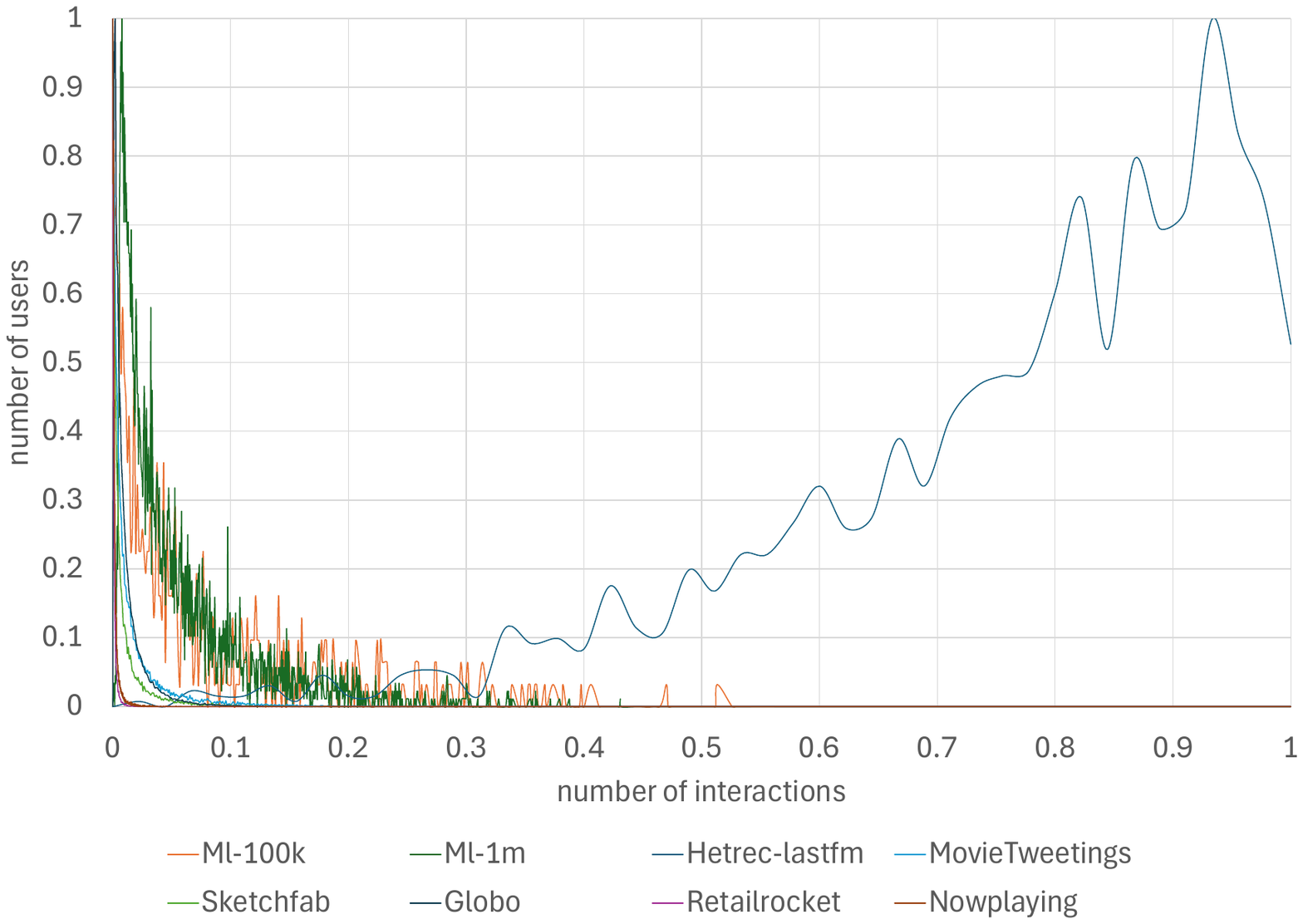}
\caption{Normalized Histogram: number of interactions to number of users.}
\label{fig:NormalizedHistogramOverview}
\end{figure}

Figure \ref{fig:NormalizedHistogramOverview} shows a normalized histogram of number of interactions with number of users as frequency over all datasets. Except for \textit{Hetrec-lastfm}, the distribution of number of users shows a similar decrease for every dataset. 
The distribution of number of users for \textit{Hetrec-lastfm} grows with number of interactions, because \textit{Hetrec-lastfm} has many users that have many interactions but a small number of users with small number of interactions. The distribution of number of users for \textit{MovieLens-100k} and \textit{MovieLens-1M} fluctuates for increasing number of interactions. 
The distribution of number of users for \textit{MovieTweetings}, \textit{Sketchfab}, \textit{Globo}, \textit{Retailrocket}, and \textit{Nowplaying} decreases gradually for increasing number of interactions.

\subsection{Experimental Pipeline}
We use four clustering approaches to create clusters on which the rest of the pipeline is executed (see Section \ref{clusterBasedAutoRecSys}).
Depending on the clustering approach we have different number of attempts to find fitting clustering parameters.
\textit{Louvain} has the least attempts with five attempts for each dataset due to five resolution values.
Both \(k\)-Means clustering approaches have seven attempts for each dataset.
\textit{Greedy Modularity} has 35 attempts due to each cluster count having five different resolutions.
We randomly split each cluster with five-fold cross validation resulting in 80\% train and 20\% test data. During the training stage \textit{Lenskit-Auto} splits the 80\% train data into 80\% train and 20\% validation data. If the training stage fails due to a lack of data, we lower the train split by 10\% and increase the validation split by 10\%.
The increasing and lowering of both splits is limited to a 50-50 split between train and validation split.
If the training still fails, we use \textit{PopScore} as a fallback recommendation algorithm.
We optimize hyperparameters and train the given recommendation algorithms with \textit{LensKit-Auto}. \textit{LensKit-Auto} performs hyperparameter optimization using \textit{Random Search}, which is limited to a search of 100 iterations. If the \textit{Random Search} exceeds four hours, it terminates the search regardless of the iterations. We evaluate the recommendation algorithms performances using \textit{nDCG@10} and \textit{Precision@10}.

\section{Results}
We structure our results in the following way. First, we present the performance of the baseline and the combined recommenders. For each clustering approach, we summarize the results in a line diagram. Each line represents a dataset and shows the \textit{nDCG@10} of the baseline and the combined recommenders. 
We conduct our experiments for \textit{nDCG@10} and \textit{Precision@10}, but only show \textit{nDCG@10} since \textit{Precision@10} shows a similar trend. We round the \textit{nDCG@10} to three positions behind the decimal point. \(K\)-Means number of interactions and \(k\)-Means item-interaction vector use the clustering parameter cluster count. \textit{Louvain} uses the clustering parameter resolution. \textit{Greedy Modularity} uses the clustering parameters cluster count and resolution, where for every cluster count the best performing cluster count over all resolution is selected. For \(k\)-Means number of interactions, \(k\)-Means item-interaction vector and \textit{Greedy Modularity} we plot cluster count on the x-axis. For \textit{Louvain} we plot resolution on the x-axis, as it generates a variable amount of clusters based on resolution. Second, we aggregate the performance of all combined recommenders over all clustering approaches. We obtain a maximum best combined recommender for each dataset. 
Third, we present the most selected recommendation algorithms per clustering approaches.
Finally, we show the average parallel runtimes of our clustering approaches compared to the baseline (base).

\begin{figure}[ht]
\centering
\includegraphics[width=0.6\columnwidth]{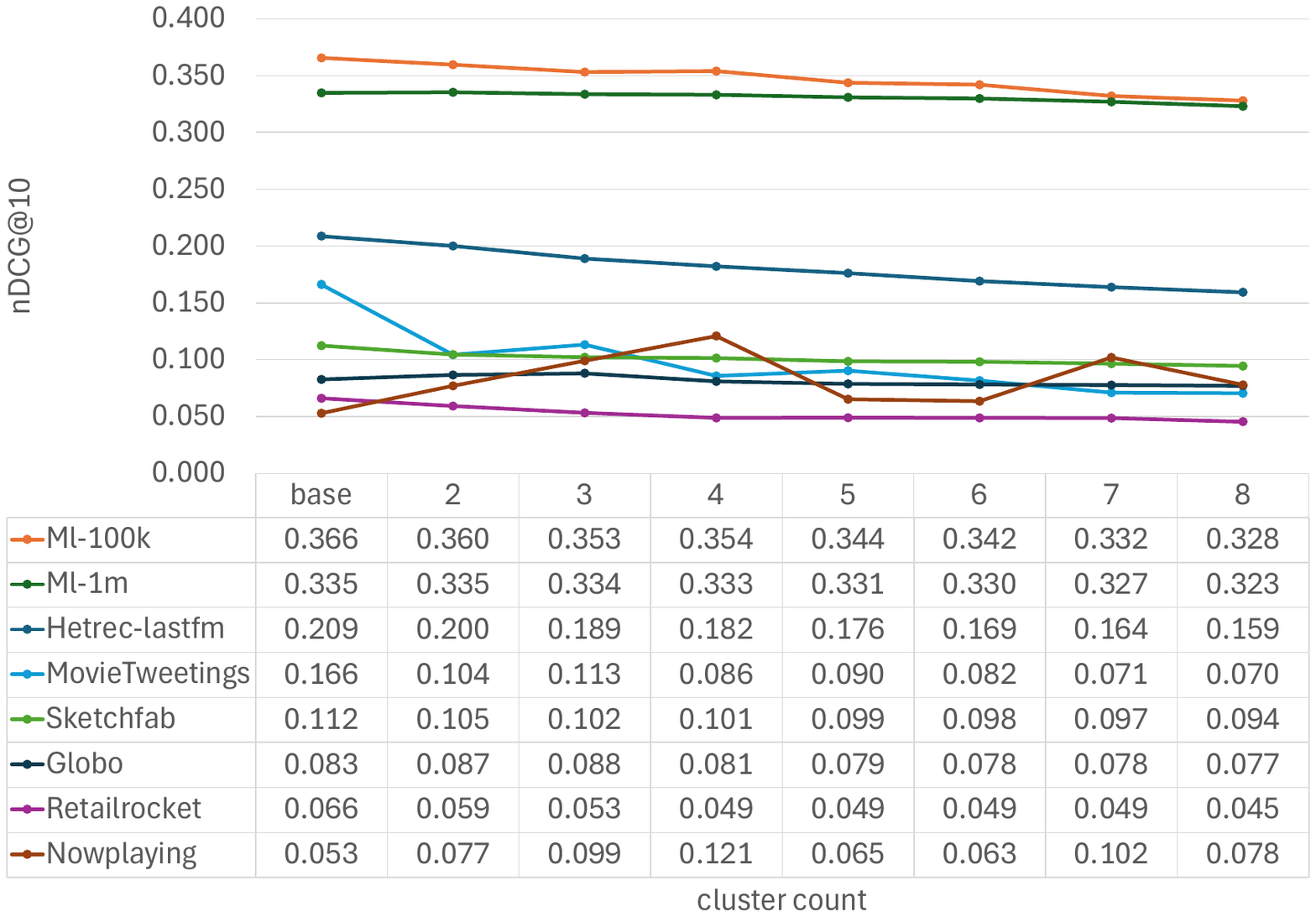}
\caption{\textit{\(k\)-Means number of interactions} with \textit{nDCG@10}.}
\label{fig:kmeansNumberOfInteractionsNdcg@10}
\end{figure}

\paragraph{Figure \ref{fig:kmeansNumberOfInteractionsNdcg@10}. \textit{\(k\)-Means number of interactions} with \textit{nDCG@10}}
\textit{\(K\)-Means number of interactions} increases the average \textit{nDCG@10} over all cluster counts for \textit{Nowplaying} by 63.07\% (0.033). \textit{Globo} has an on average decreasing \textit{nDCG@10} of 2.24\% (0.002), but the \textit{nDCG@10} increases by 6.02\% (0.005) with three clusters.
For \textit{Nowplaying}, \textit{\(k\)-Means number of interactions} increases the \textit{nDCG@10} by up to 128.30\% (0.068) with four clusters. \textit{Nowplaying} has an increase in the \textit{nDCG@10} for every cluster count. 
The \textit{nDCG@10} of \textit{Nowplaying} peaks at cluster counts four and seven, but has a plateau at five and six clusters.
\textit{MovieTweetings} has the highest \textit{nDCG@10} decrease out of all datasets. The \textit{nDCG@10} decreases by 46.99\% (0.078) on average over all cluster counts for \textit{MovieTweetings}. With an increasing cluster count the \textit{nDCG@10} for \textit{MovieTweetings} decreases.
The decrease of 57.83\% (0.096) is also the highest decrease out of all clustering approaches.
\textit{MovieLens-100k}, \textit{MovieLens-1M}, \textit{Retailrocket}, \textit{Hetrec-lastfm} and \textit{Sketchfab} have a similar decrease in \textit{nDCG@10}.

\begin{figure}[ht]
\centering
\includegraphics[width=0.6\columnwidth]{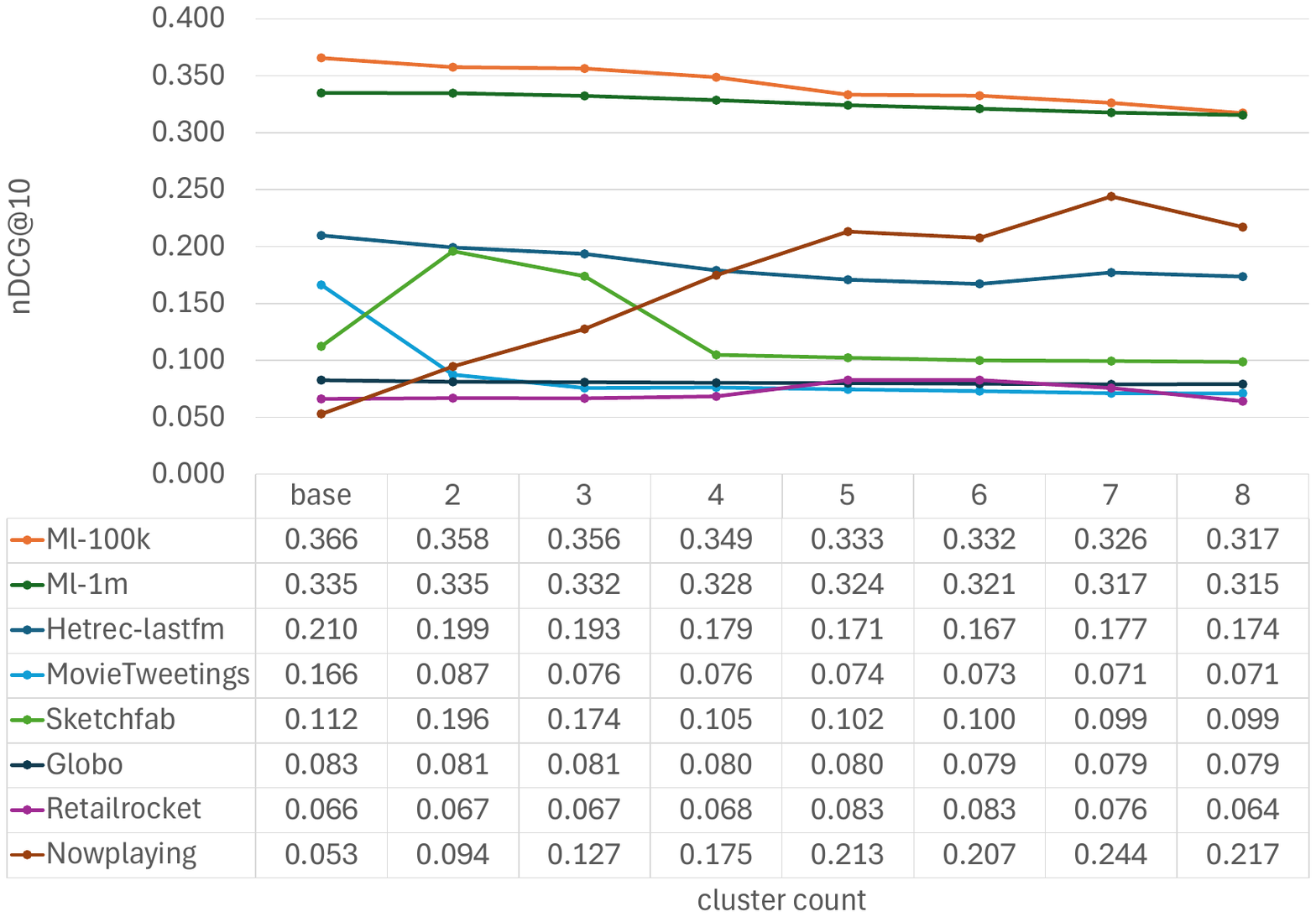}
\caption{\textit{\(k\)-Means item-interaction vector} with \textit{nDCG@10}.}
\label{fig:kmeansItemInteractionVectorNdcg@10}
\end{figure}

\paragraph{Figure \ref{fig:kmeansItemInteractionVectorNdcg@10}. \textit{\(k\)-Means item-interaction vector} with \textit{nDCG@10}}
\textit{\(K\)-Means item-interaction vector} increases the average \textit{nDCG@10} over all cluster counts for \textit{Sketchfab} by 11.61\% (0.013), \textit{Retailrocket} by 9.96\% (0.006) and \textit{Nowplaying} by 244.20\% (0.13).
For \textit{Nowplaying}, \textit{\(k\)-Means item-interaction vector} increases the \textit{nDCG@10} by up to 360.38\% (0.191) with seven clusters. \textit{Nowplaying} has an increase in the \textit{nDCG@10} for every cluster count. With an increasing cluster count the \textit{nDCG@10} for \textit{Nowplaying} increases for two to seven clusters.
\textit{\(K\)-Means item-interaction vector} increases the \textit{nDCG@10} for \textit{Sketchfab} by 75.00\% (0.084) with two clusters, and by 55.36\% (0.062) with three clusters. From four to eight clusters the \textit{nDCG@10} decreases for \textit{Sketchfab}.
\textit{MovieTweetings} has the highest \textit{nDCG@10} decrease out of all datasets. The \textit{nDCG@10} decreases by 54.58\% (0.091) on average over all cluster counts for \textit{MovieTweetings}. With an increasing cluster count the \textit{nDCG@10} for \textit{MovieTweetings} decreases.

\begin{figure}[ht]
\centering
\includegraphics[width=0.6\columnwidth]{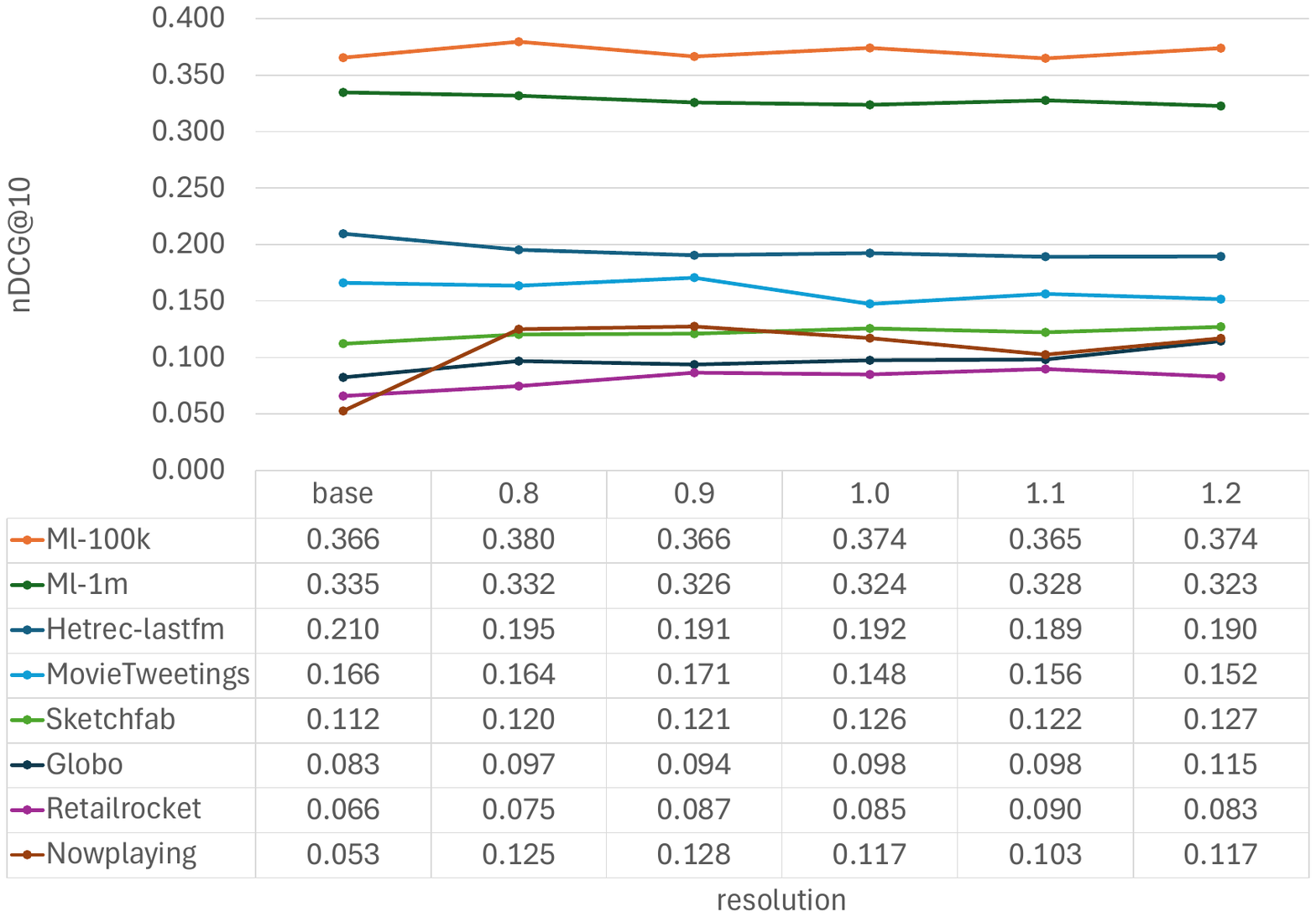}
\caption{\textit{Louvain} with \textit{nDCG@10}.}
\label{fig:LouvainCommunityDetectionNdcg@10}
\end{figure}

\begin{table}
\caption{Number of clusters for every resolution of \textit{Louvain}.}
\begin{tabular}{l|lllll}
dataset         & 0.8 & 0.9 & 1  & 1.1 & 1.2 \\ \hline
Globo           & 7   & 8   & 9  & 11  & 16  \\
Hetrec-lastfm   & 6   & 7   & 8  & 8   & 8   \\
Nowplaying      & 10  & 11  & 14 & 16  & 14  \\
Retailrocket    & 20  & 20  & 25 & 26  & 29  \\
Sketchfab       & 22  & 25  & 22 & 27  & 24  \\
MovieLens-100k  & 3   & 3   & 4  & 5   & 6   \\
MovieLens-1M    & 2   & 3   & 5  & 6   & 8   \\
MovieTweetings  & 4   & 5   & 6  & 6   & 8
\end{tabular}
\label{tab:LouvainCommunityDetectionNdcg@10ResolutionClusterCount}
\end{table}

\paragraph{Figure \ref{fig:LouvainCommunityDetectionNdcg@10}. Louvain with \textit{nDCG@10}}
\textit{Louvain} uses resolution as the clustering parameter and generates a varying number of clusters. The datasets \textit{MovieLens-100k}, \textit{MovieLens-1M}, \textit{Hetrec-lastfm}, \textit{MovieTweetings} are split into a range of three to eight clusters. The datasets \textit{Sketchfab}, \textit{Globo}, \textit{Retailrocket}, \textit{Nowplaying} are split into seven to 29 clusters (see Table \ref{tab:LouvainCommunityDetectionNdcg@10ResolutionClusterCount}).
\textit{Louvain} increases the average \textit{nDCG@10} over all cluster counts for \textit{MovieLens-100k} by 1.58\% (0.006), \textit{Retailrocket} by 27.27\% (0.018), \textit{Sketchfab} by 10.00\% (0.011), \textit{Globo} by 20.96\% (0.018) and \textit{Nowplaying} by 122.64\% (0.065).
For \textit{Nowplaying}, \textit{Louvain} increases the \textit{nDCG@10} by up to 141.51\% (0.075) with eleven clusters (0.9 resolution). \textit{Nowplaying} has an increase in the \textit{nDCG@10} for every cluster count. The datasets that show the most increase are the datasets with a low \textit{nDCG@10} for the baseline.
\textit{Louvain} increases the \textit{nDCG@10} for \textit{Globo} by an average of 16.57\% (0.014) with seven to eleven clusters (0.8 to 1.1 resolution), and by 38.55\% (0.032) with 16 clusters (1.2 resolution).
Out of all datasets, \textit{Hetrec-lastfm} has the highest average \textit{nDCG@10} decrease with 8.86\% (0.018) over all cluster counts.
\textit{Hetrec-lastfm}, \textit{MovieLens-1m}, \textit{MovieTweetings} have a similar decrease in \textit{nDCG@10}.

\begin{figure}[ht]
\centering
\includegraphics[width=0.6\columnwidth]{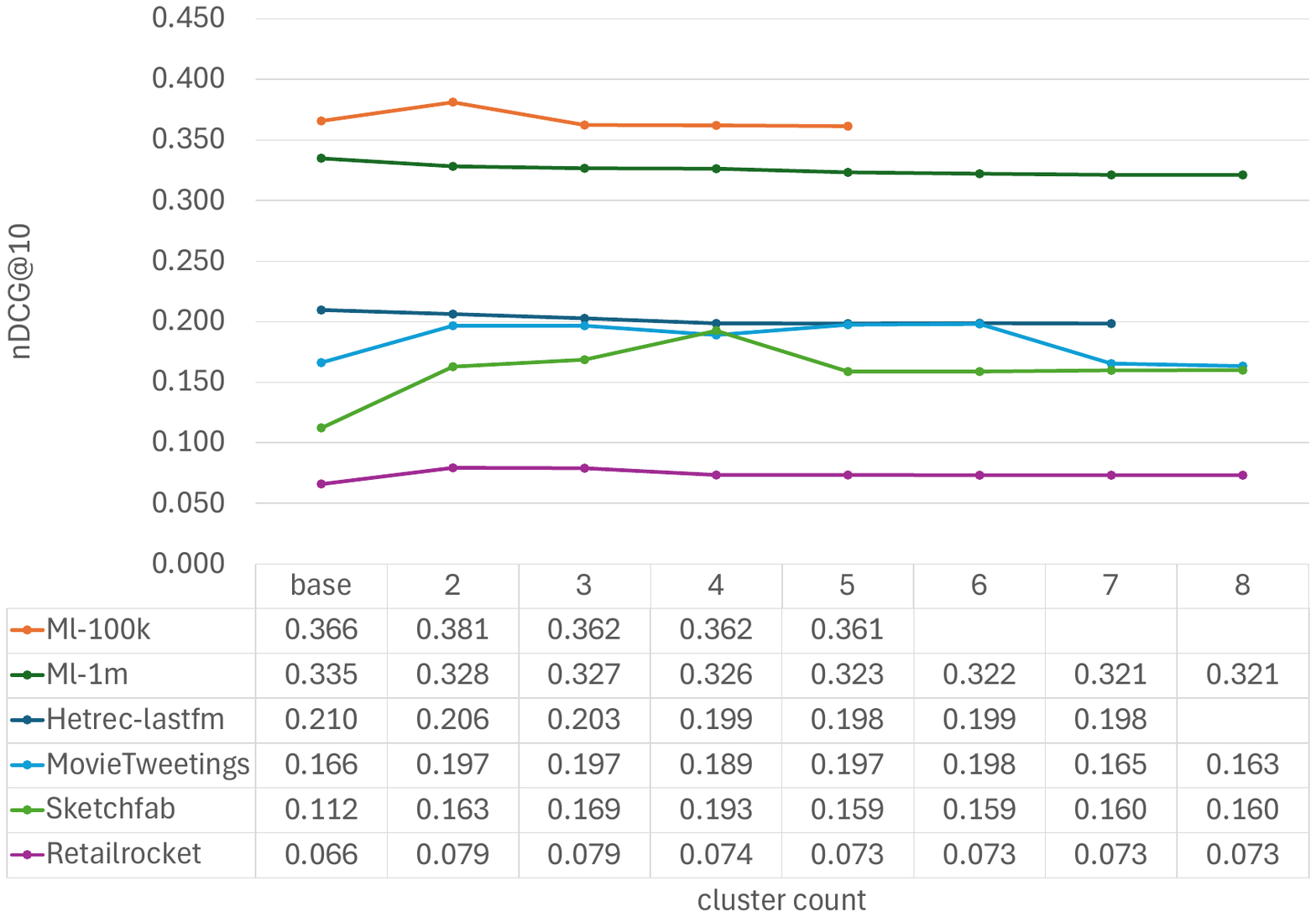}
\caption{\textit{Greedy Modularity} with \textit{nDCG@10}.}
\label{fig:GreedyModulartiyNdcg@10}
\end{figure}

\paragraph{Figure \ref{fig:GreedyModulartiyNdcg@10}. \textit{Greedy Modularity} with \textit{nDCG@10}}
\textit{Greedy Modularity} creates a maximum of five clusters for MovieLens-100k and a maximum of seven clusters for \textit{Hetrec-lastfm}. \textit{Globo} and \textit{Nowplaying} hit the time limit of six hours for the clustering stage, therefore \textit{Globo} and \textit{Nowplaying} do not have results for \textit{Greedy Modularity}.
\textit{Greedy Modularity} increases the average \textit{nDCG@10} over all cluster counts for \textit{Sketchfab} by 48.34\% (0.054) and \textit{Retailrocket} by 13.42\% (0.009).
For \textit{Sketchfab}, \textit{Greedy Modularity} increases the \textit{nDCG@10} by up to 72.32\% (0.08) with four clusters. \textit{Sketchfab} has an increase in the \textit{nDCG@10} for every cluster count. With an increasing cluster count the \textit{nDCG@10} for \textit{Sketchfab} increases for two to four clusters. For a cluster count of five to eight clusters the \textit{nDCG@10} increases on average by 42.41\% (0.047).
\textit{Greedy Modularity} increases the \textit{nDCG@10} for \textit{MovieTweetings} on average by 17.83\% (0.029) with two to six clusters. From seven to eight clusters the \textit{nDCG@10} decreases for \textit{MovieTweetings}.
For \textit{MovieLens-100k}, \textit{Greedy Modularity} increases the \textit{nDCG@10} by 4.10\% (0.016) with two clusters. From three to five clusters the \textit{nDCG@10} decreases for \textit{MovieLens-100k}.
\textit{Hetrec-lastfm} has the highest \textit{nDCG@10} decrease out of all datasets. The \textit{nDCG@10} decreases by 4.52\% (0.038) on average over all cluster counts for \textit{Hetrec-lastfm}.

\paragraph{Maximum Best Combined Recommender}
\textit{\(K\)-means number of interactions} is the best clustering approach for \textit{MovieLens-1M}, but provides no increase (0.00) in maximum \textit{nDCG@10}.
\textit{\(K\)-means item interaction vector} is the best clustering approach for \textit{Sketchfab} with a maximum \textit{nDCG@10} increase of 75\% (0.084) and \textit{Nowplaying} with a maximum \textit{nDCG@10} increase of 360\% (0.191).
\textit{Louvain} is the best clustering approach for \textit{Retailrocket} with a maximum \textit{nDCG@10} increase of 36.36\% (0.024) and \textit{Globo} with a maximum \textit{nDCG@10} increase of 38.55\% (0.032).
\textit{Greedy Modularity} is the best clustering approach for \textit{Hetrec-lastfm} with a minimum \textit{nDCG@10} decrease of 1.90\% (0.003), \textit{MovieLens-100k} with a maximum \textit{nDCG@10} increase of 4.10\% (0.016) and \textit{MovieTweetings} with a maximum \textit{nDCG@10} increase of 19.28\% (0.032).
We aggregate the maximum increases of our clustering approaches which results in an average \textit{nDCG@10} increase of 66.47\% across all datasets.

\begin{figure}[ht]
\centering
\includegraphics[width=0.6\columnwidth]{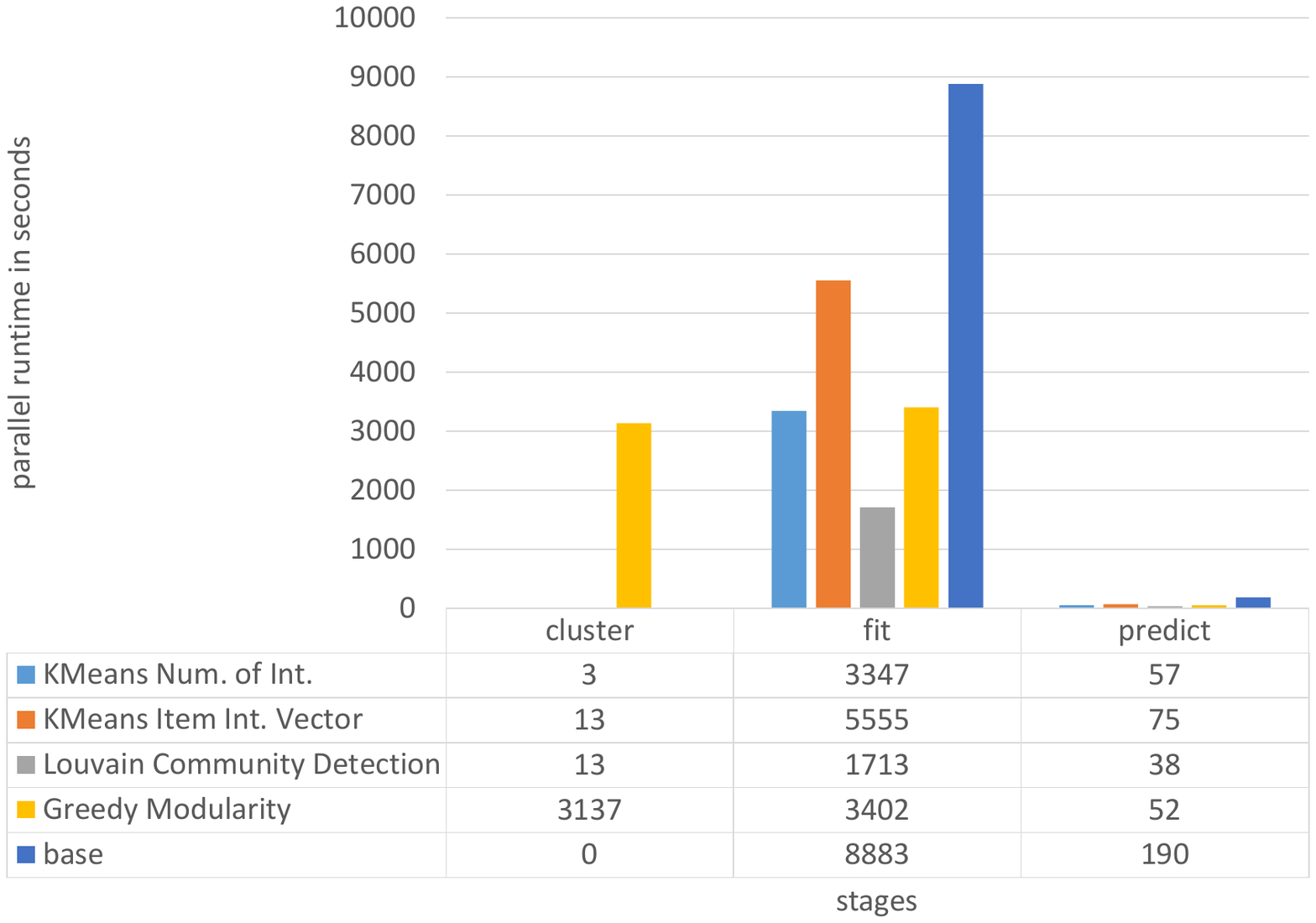}
\caption{Parallel runtime of clustering approaches and baseline.}
\label{fig:Runtime}
\end{figure}

In Figure \ref{fig:Runtime} we can see the average parallel runtime of all clustering approaches and the baseline over the stages cluster, fit and predict.
The baseline recommends over the whole dataset without the use of clusters.
The parallel runtime is the maximum runtime of a clustering approach in a given stage. 
We then average the parallel runtime over all datasets, excluding \textit{Globo} and \textit{Nowplaying}, as these two datasets hit the time limit of six hours for the clustering stage.
The biggest differences for the runtime are in the cluster and fit stage.
For the cluster stage only \textit{Greedy Modularity} shows a long runtime.
For the fit stage the baseline has the longest runtime.
\textit{\(K\)-means number of interactions} and \textit{Greedy Modularity} have a similar parallel runtime on average.
\textit{Louvain} is the clustering approach with fastest runtime overall.
In total, every clustering approach has a faster runtime than the baseline.

\begin{figure}[ht]
\centering
\includegraphics[width=0.6\columnwidth]{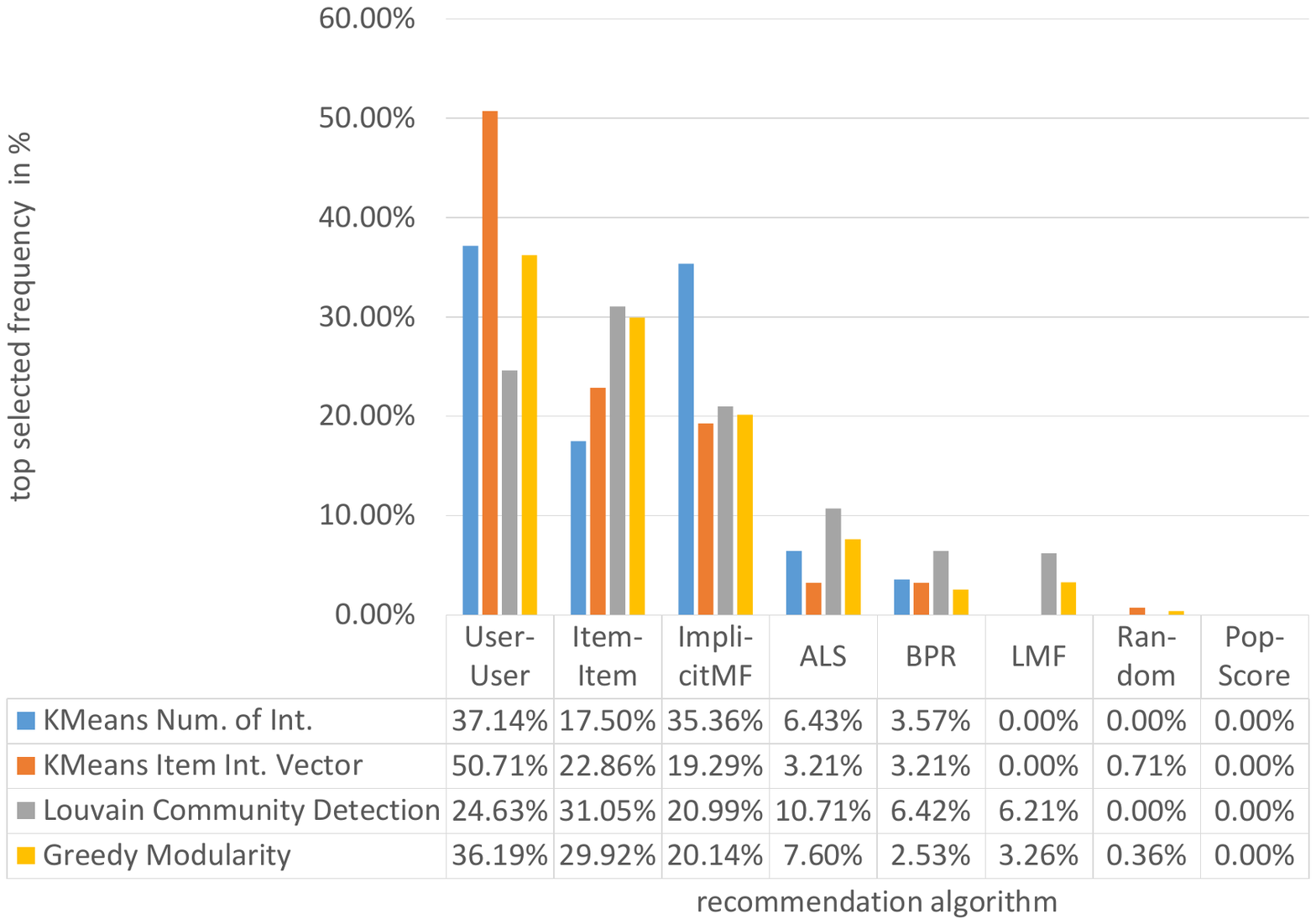}
\caption{Best performing recommendation algorithm per clustering approach.}
\label{fig:DiagramBestAlgoOverview}
\end{figure}

Figure \ref{fig:DiagramBestAlgoOverview} represents how many times a recommendation algorithm has the best \textit{nDCG@10} on each cluster per clustering approach.
The top three best performing recommendation algorithm are \textit{User-User Nearest Neighbour}, \textit{Item-Item Nearest Neighbour} and \textit{Implicit Matrix Factorization}.
\textit{User-User Nearest Neighbour} performs best for \(k\)-Means item interaction vector with 50.71\%, \(k\)-Mean number of interactions with 37.14\% and \textit{Greedy Modularity} with 36.19\%. 
\textit{Item-Item Nearest Neighbour} performs best for \textit{Louvain} with 31.05\%.
For every clustering approach a different recommendation algorithm performs best on the clusters.
The worst performing algorithms across all clusters are \textit{Random} and \textit{PopScore}. \textit{PopScore} never performs best for any cluster.

\section{Discussion}
In our results, we see that all clustering approaches have a positive influence on different datasets, except for \textit{Hetrec-lastfm}. Due to the unique distribution of \textit{Hetrec-lastfm}, we decide to exclude \textit{Hetrec-lastfm} from further discussions. All clustering approaches occur in the maximum best combined recommender. Therefore, the cluster-based algorithm selection problem cannot be solved with a single clustering approach. We conclude that we must select the optimal clustering approach to obtain the maximum best combined recommender for each dataset. We call the problem of selecting the optimal clustering approach \textit{clustering selection problem}.

Our results suggest that \textit{\(k\)-Means number of interactions} increases the recommendation performance for datasets fulfilling two criteria. First, the dataset contains significantly more items than users. Second, a significant portion of users share about the same number of interactions. We find that \textit{\(k\)-Means number of interactions} increases the average \textit{nDCG@10} across all cluster counts for \textit{Nowplaying} by 63.07\% (0.034). \textit{Nowplaying} has 47.96\% more items than users, and 38.01 average interactions per user (see Table \ref{tab:datasets}). The generated clusters separate the users that have differing number of interactions from the majority of users with a similar number of interactions. Selecting recommendation algorithms for these clusters results in the measured increase in recommendation performance.

All clustering approaches we measure for \textit{Nowplaying} increase the recommendation performance regardless of cluster count and resolution. A reason for \textit{Nowplaying} to benefit from all clustering approaches can be the sheer number of users and items and average number of interactions of 38.01 per user. Additionally, \textit{Nowplaying} contains 47.96\% more items than users. We assume that clustering in general has a positive influence for large datasets.

The smallest recommendation performance increases are on \textit{MovieLens-1m}, \textit{MovieLens-100k} and \textit{MovieTweetings}. These three datasets are the only explicit feedback datasets. We convert the explicit feedback into implicit feedback by counting every rating as an interaction. This conversion could distort the meta-information used by our clustering approaches. For example, user A with ten 5 star ratings and user B with ten 1 star ratings are both converted into users with ten interactions. Our clustering approaches would cluster user A and B into the same cluster, although their interactions are different.

\textit{MovieLens-1m} and \textit{MovieLens-100k} have a minimum number of 20 interactions per user, and not a minimum number of five interactions per user like the other datasets. A different threshold for pruning can decrease the influence of clustering. A big portion of users, on which clustering could improve recommendation performance, is missing. The big portion of missing users presents a missed opportunity for clustering to separate users.
Additionally, clustering can only increase the performance of recommendation algorithms when the clusters use different best recommendation algorithms. If the same recommendation algorithm is used for all clusters, the performance remains the same at best. For example, \textit{User-User Nearest Neighbour} is the single best recommendation algorithm and is selected for most clusters across all clustering approaches for \textit{MovieLens-1m}.

\textit{Random} algorithm is unexpectedly selected in 0.71\% (2) of cases for \textit{\(k\)-Means item interaction vector} and in 0.36\% (3) of cases for \textit{Greedy Modularity} (see Figure \ref{fig:DiagramBestAlgoOverview}). Without clustering, the random algorithm has an average \textit{nDCG@10} of 0.003, which is always below every other recommendation algorithm. The reason for the unexpected recommendation performance of the \textit{Random} algorithm is that clustering can result in clusters with few users who have many interactions. For those few users the \textit{Random} recommendation algorithm can recommend the correct items by pure chance. Therefore, \textit{Random} recommendation algorithm is in five cases the best.

We see multiple reasons for \textit{Louvain} being the best clustering approach. \textit{Louvain} is the overall best clustering approach increasing the recommendation performance on six out of eight datasets. \textit{Louvain} scales well for larger datasets, because \textit{Louvain} performs clustering based on resolution resulting in a variable cluster count. \textit{Louvain} already performs well with default resolution increasing the \textit{nDCG@10} on average by 19.95\% across all datasets. \textit{Louvain} has on average the smallest parallel runtime among all clustering approaches.

Clustering can improve hyperparameter optimization. For large datasets, the recommendation algorithms have a long runtime and may not complete all 100 iterations of random search hitting the time limit. Clusters are smaller and recommendation algorithms can run up to 100 iterations of random search and therefore can find better hyperparameters, resulting in better recommendation performance. However, smaller clusters do not guarantee a performance increase. Otherwise our combined recommenders regardless of the clustering approach should increase the recommendation performance. The choice of clustering approach is the decisive factor.

Clustering speeds up the parallel runtime of hyperparameter optimization (see Figure \ref{fig:Runtime}). The parallel runtime is the maximum runtime of the hyperparameter optimization of all clusters. Due to clustering, we have less users to optimize for in each optimization. While the number of runs increases by one for each cluster, each run is fully parallelizable.
If we parallelize every stage our total parallel runtime is always smaller than the baseline (see Figure \ref{fig:Runtime}). Due to clustering we see a considerable decrease in parallel runtime for the fit stage. The decrease of parallel runtime can be accounted to less users in a cluster which need to be fit. During the fit stage, \textit{Greedy Modularity} has an average parallel runtime compared to the other clustering approaches. However, \textit{Greedy Modularity} is the only clustering approach with a noticeable parallel runtime for the cluster stage. Summing up the parallel runtime of the cluster and fit stage, \textit{Greedy Modularity} is the clustering approach with the longest parallel runtime.

\section{Conclusion}
In our research question we ask how clustering influences the performance of recommendation algorithms. We demonstrate that cluster-based algorithm selection can yield significantly better results than algorithm selection without clustering. For five out of eight datasets we see a drastic increase in \textit{nDCG@10} that ranges between 19.28\% (0.032) and 360.38\% (0.191). We demonstrate that clustering is an effective technique for optimizing recommendation algorithm performance. For \textit{MovieLens-100k} and \textit{MovieLens-1m} the single best recommender is at the same time the best choice for most clusters resulting in performance increases of 4.10\% (0.016) and 0.00\% with clustering. We might see an increase for a larger selection of recommendation algorithms. 

We derive the clustering selection problem as the problem of selecting the optimal clustering approach that yields an increase in performance for a given dataset. Our work shows, there is no single best clustering approach that fits all datasets. We suggest to test a wide variety of clustering approaches and recommendation algorithms.

Clustering has been overlooked as a valid approach to increase performance of recommendation algorithms. We hope that our work changes the perception of the community on clustering for recommender systems. As seen in our experiments, we confirm that certain combinations of datasets, clustering approaches and recommendation algorithms show no performance increase. We identified a downward trend and plateauing for over half of our combinations. This appears to be in line with related work. However, as seen with the maximum best combined recommender, we only need to find a single combination to drastically improve performance.

\section{Future Work and Limitations}
Future work could consider more clustering approaches and meta-information to further improve cluster-based algorithm selection.
In our work, we use four clustering approaches. Each clustering approach considers one specific meta-information derived from the datasets, e.g. number of interactions, even though the datasets may contain additional meta-information for the clustering approaches. We do not know which meta-information enables our clustering approaches to increase recommendation performance. The lack of performance increasing meta-information limits the capabilities of our clustering approaches to perform cluster-based algorithm selection.

Future work should further analyze the structure of the datasets and the relation of the structure to the resulting recommendation performance. We hope the analysis of the dataset structure provides insights to better select the best clustering approach and even clustering parameters for the datasets.
The datasets we use in our work vary in structure such as number of users, number of items, number of interactions, sparsity, etc. Cluster-based algorithm selection uses parts of the structural meta-information to increase recommendation performance. The recommendation performances achieved by our clustering approaches varies drastically across different datasets. We can not find a conclusive connection between the structural meta-information and the recommendation performances achieved by our clustering approaches yet. 

Future work could research the correlation between clustering-quality measures, e.g. silhouette score, and the recommendation performance of the maximum best combined recommender to reduce training overhead.
We currently determine the influence of clustering on the recommendation performance after training and predicting. To determine the influence of clustering on the recommendation performance, we train recommendation algorithms on each cluster for each set of clusters. We use the recommendation performance of the combined recommenders on each set of clusters to select the maximum best combined recommender. This "trial and error" method leads to a lot of trained recommendation algorithms which will never be used. Knowing which clustering approach and parameters perform well before training and predicting saves a lot of time. We require additional metrics to estimate the resulting recommendation performance of the clustering approach and clustering parameters before training.

\begin{acks}
We thank Tobias Vente for supervising our work.
The results reported in this research were computed using the OMNI cluster of the University of Siegen.
\end{acks}

\bibliographystyle{ACM-Reference-Format}
\bibliography{references}


\begin{thebibliography}{45}


\ifx \showCODEN    \undefined \def \showCODEN     #1{\unskip}     \fi
\ifx \showDOI      \undefined \def \showDOI       #1{#1}\fi
\ifx \showISBNx    \undefined \def \showISBNx     #1{\unskip}     \fi
\ifx \showISBNxiii \undefined \def \showISBNxiii  #1{\unskip}     \fi
\ifx \showISSN     \undefined \def \showISSN      #1{\unskip}     \fi
\ifx \showLCCN     \undefined \def \showLCCN      #1{\unskip}     \fi
\ifx \shownote     \undefined \def \shownote      #1{#1}          \fi
\ifx \showarticletitle \undefined \def \showarticletitle #1{#1}   \fi
\ifx \showURL      \undefined \def \showURL       {\relax}        \fi
\providecommand\bibfield[2]{#2}
\providecommand\bibinfo[2]{#2}
\providecommand\natexlab[1]{#1}
\providecommand\showeprint[2][]{arXiv:#2}

\bibitem[Beel et~al\mbox{.}(2016)]%
        {beel2016towards}
\bibfield{author}{\bibinfo{person}{Joeran Beel}, \bibinfo{person}{Corinna Breitinger}, \bibinfo{person}{Stefan Langer}, \bibinfo{person}{Andreas Lommatzsch}, {and} \bibinfo{person}{Bela Gipp}.} \bibinfo{year}{2016}\natexlab{}.
\newblock \showarticletitle{Towards reproducibility in recommender-systems research}.
\newblock \bibinfo{journal}{\emph{User Modeling and User-Adapted Interaction}} \bibinfo{volume}{26}, \bibinfo{number}{1} (\bibinfo{date}{mar} \bibinfo{year}{2016}), \bibinfo{pages}{69–101}.
\newblock
\showISSN{0924-1868}
\urldef\tempurl%
\url{https://doi.org/10.1007/s11257-016-9174-x}
\showDOI{\tempurl}


\bibitem[Beel and Brunel(2019)]%
        {beelPruning}
\bibfield{author}{\bibinfo{person}{Joeran Beel} {and} \bibinfo{person}{Victor Brunel}.} \bibinfo{year}{2019}\natexlab{}.
\newblock \showarticletitle{Data pruning in recommender systems research: Best-practice or malpractice}.
\newblock \bibinfo{journal}{\emph{ACM RecSys}} (\bibinfo{year}{2019}).
\newblock


\bibitem[Bellogin and Parapar(2012)]%
        {10.1145/2365952.2365997}
\bibfield{author}{\bibinfo{person}{Alejandro Bellogin} {and} \bibinfo{person}{Javier Parapar}.} \bibinfo{year}{2012}\natexlab{}.
\newblock \showarticletitle{Using graph partitioning techniques for neighbour selection in user-based collaborative filtering}. In \bibinfo{booktitle}{\emph{Proceedings of the Sixth ACM Conference on Recommender Systems}} (Dublin, Ireland) \emph{(\bibinfo{series}{RecSys '12})}. \bibinfo{publisher}{Association for Computing Machinery}, \bibinfo{address}{New York, NY, USA}, \bibinfo{pages}{213–216}.
\newblock
\showISBNx{9781450312707}
\urldef\tempurl%
\url{https://doi.org/10.1145/2365952.2365997}
\showDOI{\tempurl}


\bibitem[Blondel et~al\mbox{.}(2008)]%
        {louvain}
\bibfield{author}{\bibinfo{person}{Vincent~D Blondel}, \bibinfo{person}{Jean-Loup Guillaume}, \bibinfo{person}{Renaud Lambiotte}, {and} \bibinfo{person}{Etienne Lefebvre}.} \bibinfo{year}{2008}\natexlab{}.
\newblock \showarticletitle{Fast unfolding of communities in large networks}.
\newblock \bibinfo{journal}{\emph{Journal of Statistical Mechanics: Theory and Experiment}} \bibinfo{volume}{2008}, \bibinfo{number}{10} (\bibinfo{date}{oct} \bibinfo{year}{2008}), \bibinfo{pages}{P10008}.
\newblock
\urldef\tempurl%
\url{https://doi.org/10.1088/1742-5468/2008/10/P10008}
\showDOI{\tempurl}


\bibitem[Cantador et~al\mbox{.}(2011)]%
        {10.1145/2043932.2044016}
\bibfield{author}{\bibinfo{person}{Ivan Cantador}, \bibinfo{person}{Peter Brusilovsky}, {and} \bibinfo{person}{Tsvi Kuflik}.} \bibinfo{year}{2011}\natexlab{}.
\newblock \showarticletitle{Second workshop on information heterogeneity and fusion in recommender systems (HetRec2011)}. In \bibinfo{booktitle}{\emph{Proceedings of the Fifth ACM Conference on Recommender Systems}} (Chicago, Illinois, USA) \emph{(\bibinfo{series}{RecSys '11})}. \bibinfo{publisher}{Association for Computing Machinery}, \bibinfo{address}{New York, NY, USA}, \bibinfo{pages}{387–388}.
\newblock
\showISBNx{9781450306836}
\urldef\tempurl%
\url{https://doi.org/10.1145/2043932.2044016}
\showDOI{\tempurl}


\bibitem[Clauset et~al\mbox{.}(2004)]%
        {greedy_modularity}
\bibfield{author}{\bibinfo{person}{Aaron Clauset}, \bibinfo{person}{M.~E.~J. Newman}, {and} \bibinfo{person}{Cristopher Moore}.} \bibinfo{year}{2004}\natexlab{}.
\newblock \showarticletitle{Finding community structure in very large networks}.
\newblock \bibinfo{journal}{\emph{Phys. Rev. E}}  \bibinfo{volume}{70} (\bibinfo{date}{Dec} \bibinfo{year}{2004}), \bibinfo{pages}{066111}.
\newblock
Issue 6.
\urldef\tempurl%
\url{https://doi.org/10.1103/PhysRevE.70.066111}
\showDOI{\tempurl}


\bibitem[Collins et~al\mbox{.}(2018)]%
        {collins2018novel}
\bibfield{author}{\bibinfo{person}{Andrew Collins}, \bibinfo{person}{Dominika Tkaczyk}, {and} \bibinfo{person}{Joeran Beel}.} \bibinfo{year}{2018}\natexlab{}.
\newblock \showarticletitle{A Novel Approach to Recommendation Algorithm Selection using Meta-Learning.}. In \bibinfo{booktitle}{\emph{AICS}}. \bibinfo{pages}{210--219}.
\newblock


\bibitem[Cunha et~al\mbox{.}(2016)]%
        {10.1007/978-3-319-46227-1_25}
\bibfield{author}{\bibinfo{person}{Tiago Cunha}, \bibinfo{person}{Carlos Soares}, {and} \bibinfo{person}{Andr{\'e} C. P. L.~F. de Carvalho}.} \bibinfo{year}{2016}\natexlab{}.
\newblock \showarticletitle{Selecting Collaborative Filtering Algorithms Using Metalearning}. In \bibinfo{booktitle}{\emph{Machine Learning and Knowledge Discovery in Databases}}, \bibfield{editor}{\bibinfo{person}{Paolo Frasconi}, \bibinfo{person}{Niels Landwehr}, \bibinfo{person}{Giuseppe Manco}, {and} \bibinfo{person}{Jilles Vreeken}} (Eds.). \bibinfo{publisher}{Springer International Publishing}, \bibinfo{address}{Cham}, \bibinfo{pages}{393--409}.
\newblock
\showISBNx{978-3-319-46227-1}


\bibitem[Cunha et~al\mbox{.}(2018)]%
        {10.1145/3240323.3240378}
\bibfield{author}{\bibinfo{person}{Tiago Cunha}, \bibinfo{person}{Carlos Soares}, {and} \bibinfo{person}{Andr\'{e} C. P. L.~F. de Carvalho}.} \bibinfo{year}{2018}\natexlab{}.
\newblock \showarticletitle{CF4CF: recommending collaborative filtering algorithms using collaborative filtering}. In \bibinfo{booktitle}{\emph{Proceedings of the 12th ACM Conference on Recommender Systems}} (Vancouver, British Columbia, Canada) \emph{(\bibinfo{series}{RecSys '18})}. \bibinfo{publisher}{Association for Computing Machinery}, \bibinfo{address}{New York, NY, USA}, \bibinfo{pages}{357–361}.
\newblock
\showISBNx{9781450359016}
\urldef\tempurl%
\url{https://doi.org/10.1145/3240323.3240378}
\showDOI{\tempurl}


\bibitem[Da~Costa and Manzato(2014)]%
        {6984809}
\bibfield{author}{\bibinfo{person}{Arthur~F. Da~Costa} {and} \bibinfo{person}{Marcelo~Garcia Manzato}.} \bibinfo{year}{2014}\natexlab{}.
\newblock \showarticletitle{Multimodal Interactions in Recommender Systems: An Ensembling Approach}. In \bibinfo{booktitle}{\emph{2014 Brazilian Conference on Intelligent Systems}}. \bibinfo{pages}{67--72}.
\newblock
\urldef\tempurl%
\url{https://doi.org/10.1109/BRACIS.2014.23}
\showDOI{\tempurl}


\bibitem[da~Costa~Fortes and Manzato(2014)]%
        {10.1145/2664551.2664556}
\bibfield{author}{\bibinfo{person}{Arthur da Costa~Fortes} {and} \bibinfo{person}{Marcelo~Garcia Manzato}.} \bibinfo{year}{2014}\natexlab{}.
\newblock \showarticletitle{Ensemble Learning in Recommender Systems: Combining Multiple User Interactions for Ranking Personalization}. In \bibinfo{booktitle}{\emph{Proceedings of the 20th Brazilian Symposium on Multimedia and the Web}} (Jo\~{a}o Pessoa, Brazil) \emph{(\bibinfo{series}{WebMedia '14})}. \bibinfo{publisher}{Association for Computing Machinery}, \bibinfo{address}{New York, NY, USA}, \bibinfo{pages}{47–54}.
\newblock
\showISBNx{9781450332309}
\urldef\tempurl%
\url{https://doi.org/10.1145/2664551.2664556}
\showDOI{\tempurl}


\bibitem[de~Souza Pereira~Moreira et~al\mbox{.}(2018)]%
        {10.1145/3270323.3270328}
\bibfield{author}{\bibinfo{person}{Gabriel de Souza Pereira~Moreira}, \bibinfo{person}{Felipe Ferreira}, {and} \bibinfo{person}{Adilson~Marques da Cunha}.} \bibinfo{year}{2018}\natexlab{}.
\newblock \showarticletitle{News Session-Based Recommendations using Deep Neural Networks}. In \bibinfo{booktitle}{\emph{Proceedings of the 3rd Workshop on Deep Learning for Recommender Systems}} (Vancouver, BC, Canada) \emph{(\bibinfo{series}{DLRS 2018})}. \bibinfo{publisher}{Association for Computing Machinery}, \bibinfo{address}{New York, NY, USA}, \bibinfo{pages}{15–23}.
\newblock
\showISBNx{9781450366175}
\urldef\tempurl%
\url{https://doi.org/10.1145/3270323.3270328}
\showDOI{\tempurl}


\bibitem[Dooms(2021)]%
        {MovieTweetings}
\bibfield{author}{\bibinfo{person}{Simon Dooms}.} \bibinfo{year}{2021}\natexlab{}.
\newblock \bibinfo{title}{A Live Movie Rating Dataset Collected From Twitter}.
\newblock
\newblock
\urldef\tempurl%
\url{https://github.com/sidooms/MovieTweetings}
\showURL{%
\tempurl}


\bibitem[Edenhofer et~al\mbox{.}(2019)]%
        {edenhofer2019augmenting}
\bibfield{author}{\bibinfo{person}{Gordian Edenhofer}, \bibinfo{person}{Andrew Collins}, \bibinfo{person}{Akiko Aizawa}, {and} \bibinfo{person}{Joeran Beel}.} \bibinfo{year}{2019}\natexlab{}.
\newblock \showarticletitle{Augmenting the DonorsChoose. org Corpus for Meta-Learning.}. In \bibinfo{booktitle}{\emph{AMIR@ ECIR}}. \bibinfo{pages}{32--38}.
\newblock


\bibitem[Ekstrand and Riedl(2012)]%
        {10.1145/2365952.2366002}
\bibfield{author}{\bibinfo{person}{Michael Ekstrand} {and} \bibinfo{person}{John Riedl}.} \bibinfo{year}{2012}\natexlab{}.
\newblock \showarticletitle{When recommenders fail: predicting recommender failure for algorithm selection and combination}. In \bibinfo{booktitle}{\emph{Proceedings of the Sixth ACM Conference on Recommender Systems}} (Dublin, Ireland) \emph{(\bibinfo{series}{RecSys '12})}. \bibinfo{publisher}{Association for Computing Machinery}, \bibinfo{address}{New York, NY, USA}, \bibinfo{pages}{233–236}.
\newblock
\showISBNx{9781450312707}
\urldef\tempurl%
\url{https://doi.org/10.1145/2365952.2366002}
\showDOI{\tempurl}


\bibitem[Ekstrand(2020)]%
        {10.1145/3340531.3412778}
\bibfield{author}{\bibinfo{person}{Michael~D. Ekstrand}.} \bibinfo{year}{2020}\natexlab{}.
\newblock \showarticletitle{LensKit for Python: Next-Generation Software for Recommender Systems Experiments}. In \bibinfo{booktitle}{\emph{Proceedings of the 29th ACM International Conference on Information \& Knowledge Management}} (Virtual Event, Ireland) \emph{(\bibinfo{series}{CIKM '20})}. \bibinfo{publisher}{Association for Computing Machinery}, \bibinfo{address}{New York, NY, USA}, \bibinfo{pages}{2999–3006}.
\newblock
\showISBNx{9781450368599}
\urldef\tempurl%
\url{https://doi.org/10.1145/3340531.3412778}
\showDOI{\tempurl}


\bibitem[Forouzandeh et~al\mbox{.}(2021)]%
        {forouzandeh2021presentation}
\bibfield{author}{\bibinfo{person}{Saman Forouzandeh}, \bibinfo{person}{Kamal Berahmand}, {and} \bibinfo{person}{Mehrdad Rostami}.} \bibinfo{year}{2021}\natexlab{}.
\newblock \showarticletitle{Presentation of a recommender system with ensemble learning and graph embedding: a case on MovieLens}.
\newblock \bibinfo{journal}{\emph{Multimedia Tools and Applications}} \bibinfo{volume}{80}, \bibinfo{number}{5} (\bibinfo{year}{2021}), \bibinfo{pages}{7805--7832}.
\newblock
\urldef\tempurl%
\url{https://doi.org/10.1007/s11042-020-09949-5}
\showDOI{\tempurl}


\bibitem[Frederickson(2023)]%
        {githubImplicit}
\bibfield{author}{\bibinfo{person}{Ben Frederickson}.} \bibinfo{year}{2023}\natexlab{}.
\newblock \bibinfo{title}{Fast Python Collaborative Filtering for Implicit Feedback Datasets}.
\newblock
\newblock
\urldef\tempurl%
\url{https://github.com/benfred/implicit}
\showURL{%
\tempurl}


\bibitem[Guo et~al\mbox{.}(2015)]%
        {GUO201514}
\bibfield{author}{\bibinfo{person}{Guibing Guo}, \bibinfo{person}{Jie Zhang}, {and} \bibinfo{person}{Neil Yorke-Smith}.} \bibinfo{year}{2015}\natexlab{}.
\newblock \showarticletitle{Leveraging multiviews of trust and similarity to enhance clustering-based recommender systems}.
\newblock \bibinfo{journal}{\emph{Knowledge-Based Systems}}  \bibinfo{volume}{74} (\bibinfo{year}{2015}), \bibinfo{pages}{14--27}.
\newblock
\showISSN{0950-7051}
\urldef\tempurl%
\url{https://doi.org/10.1016/j.knosys.2014.10.016}
\showDOI{\tempurl}


\bibitem[Guo et~al\mbox{.}(2022)]%
        {9216015}
\bibfield{author}{\bibinfo{person}{Qingyu Guo}, \bibinfo{person}{Fuzhen Zhuang}, \bibinfo{person}{Chuan Qin}, \bibinfo{person}{Hengshu Zhu}, \bibinfo{person}{Xing Xie}, \bibinfo{person}{Hui Xiong}, {and} \bibinfo{person}{Qing He}.} \bibinfo{year}{2022}\natexlab{}.
\newblock \showarticletitle{A Survey on Knowledge Graph-Based Recommender Systems}.
\newblock \bibinfo{journal}{\emph{IEEE Transactions on Knowledge and Data Engineering}} \bibinfo{volume}{34}, \bibinfo{number}{8} (\bibinfo{date}{Aug} \bibinfo{year}{2022}), \bibinfo{pages}{3549--3568}.
\newblock
\showISSN{1558-2191}
\urldef\tempurl%
\url{https://doi.org/10.1109/TKDE.2020.3028705}
\showDOI{\tempurl}


\bibitem[Harper and Konstan(2015)]%
        {10.1145/2827872}
\bibfield{author}{\bibinfo{person}{F.~Maxwell Harper} {and} \bibinfo{person}{Joseph~A. Konstan}.} \bibinfo{year}{2015}\natexlab{}.
\newblock \showarticletitle{The MovieLens Datasets: History and Context}.
\newblock \bibinfo{journal}{\emph{ACM Trans. Interact. Intell. Syst.}} \bibinfo{volume}{5}, \bibinfo{number}{4}, Article \bibinfo{articleno}{19} (\bibinfo{date}{dec} \bibinfo{year}{2015}), \bibinfo{numpages}{19}~pages.
\newblock
\showISSN{2160-6455}
\urldef\tempurl%
\url{https://doi.org/10.1145/2827872}
\showDOI{\tempurl}


\bibitem[Jahrer et~al\mbox{.}(2010)]%
        {10.1145/1835804.1835893}
\bibfield{author}{\bibinfo{person}{Michael Jahrer}, \bibinfo{person}{Andreas T\"{o}scher}, {and} \bibinfo{person}{Robert Legenstein}.} \bibinfo{year}{2010}\natexlab{}.
\newblock \showarticletitle{Combining predictions for accurate recommender systems}. In \bibinfo{booktitle}{\emph{Proceedings of the 16th ACM SIGKDD International Conference on Knowledge Discovery and Data Mining}} (Washington, DC, USA) \emph{(\bibinfo{series}{KDD '10})}. \bibinfo{publisher}{Association for Computing Machinery}, \bibinfo{address}{New York, NY, USA}, \bibinfo{pages}{693–702}.
\newblock
\showISBNx{9781450300551}
\urldef\tempurl%
\url{https://doi.org/10.1145/1835804.1835893}
\showDOI{\tempurl}


\bibitem[Jalili et~al\mbox{.}(2018)]%
        {8550639}
\bibfield{author}{\bibinfo{person}{Mahdi Jalili}, \bibinfo{person}{Sajad Ahmadian}, \bibinfo{person}{Maliheh Izadi}, \bibinfo{person}{Parham Moradi}, {and} \bibinfo{person}{Mostafa Salehi}.} \bibinfo{year}{2018}\natexlab{}.
\newblock \showarticletitle{Evaluating Collaborative Filtering Recommender Algorithms: A Survey}.
\newblock \bibinfo{journal}{\emph{IEEE Access}}  \bibinfo{volume}{6} (\bibinfo{year}{2018}), \bibinfo{pages}{74003--74024}.
\newblock
\urldef\tempurl%
\url{https://doi.org/10.1109/ACCESS.2018.2883742}
\showDOI{\tempurl}


\bibitem[Li and Kim(2003)]%
        {1241167}
\bibfield{author}{\bibinfo{person}{Qing Li} {and} \bibinfo{person}{Byeong~Man Kim}.} \bibinfo{year}{2003}\natexlab{}.
\newblock \showarticletitle{Clustering approach for hybrid recommender system}. In \bibinfo{booktitle}{\emph{Proceedings IEEE/WIC International Conference on Web Intelligence (WI 2003)}}. \bibinfo{pages}{33--38}.
\newblock
\urldef\tempurl%
\url{https://doi.org/10.1109/WI.2003.1241167}
\showDOI{\tempurl}


\bibitem[MacQueen et~al\mbox{.}(1967)]%
        {kmean}
\bibfield{author}{\bibinfo{person}{James MacQueen} {et~al\mbox{.}}} \bibinfo{year}{1967}\natexlab{}.
\newblock \showarticletitle{Some methods for classification and analysis of multivariate observations}. In \bibinfo{booktitle}{\emph{Proceedings of the fifth Berkeley symposium on mathematical statistics and probability}}, Vol.~\bibinfo{volume}{1}. Oakland, CA, USA, \bibinfo{pages}{281--297}.
\newblock


\bibitem[Matuszyk and Spiliopoulou(2014)]%
        {10.1145/2611040.2611054}
\bibfield{author}{\bibinfo{person}{Pawel Matuszyk} {and} \bibinfo{person}{Myra Spiliopoulou}.} \bibinfo{year}{2014}\natexlab{}.
\newblock \showarticletitle{Predicting the Performance of Collaborative Filtering Algorithms}. In \bibinfo{booktitle}{\emph{Proceedings of the 4th International Conference on Web Intelligence, Mining and Semantics (WIMS14)}} (Thessaloniki, Greece) \emph{(\bibinfo{series}{WIMS '14})}. \bibinfo{publisher}{Association for Computing Machinery}, \bibinfo{address}{New York, NY, USA}, Article \bibinfo{articleno}{38}, \bibinfo{numpages}{6}~pages.
\newblock
\showISBNx{9781450325387}
\urldef\tempurl%
\url{https://doi.org/10.1145/2611040.2611054}
\showDOI{\tempurl}


\bibitem[Meltsov(2019)]%
        {meltsov2019utilizing}
\bibfield{author}{\bibinfo{person}{Vasily Meltsov}.} \bibinfo{year}{2019}\natexlab{}.
\newblock \showarticletitle{Utilizing Metadata to Select a Recommendation Algorithm for a User or an Item}.
\newblock  (\bibinfo{year}{2019}).
\newblock


\bibitem[Moreira et~al\mbox{.}(2019)]%
        {8908688}
\bibfield{author}{\bibinfo{person}{Gabriel De Souza~P. Moreira}, \bibinfo{person}{Dietmar Jannach}, {and} \bibinfo{person}{Adilson Marques~Da Cunha}.} \bibinfo{year}{2019}\natexlab{}.
\newblock \showarticletitle{Contextual Hybrid Session-Based News Recommendation With Recurrent Neural Networks}.
\newblock \bibinfo{journal}{\emph{IEEE Access}}  \bibinfo{volume}{7} (\bibinfo{year}{2019}), \bibinfo{pages}{169185--169203}.
\newblock
\urldef\tempurl%
\url{https://doi.org/10.1109/ACCESS.2019.2954957}
\showDOI{\tempurl}


\bibitem[netflix.com(2023)]%
        {netflix_report}
\bibfield{author}{\bibinfo{person}{netflix.com}.} \bibinfo{year}{2023}\natexlab{}.
\newblock \bibinfo{booktitle}{\emph{What We Watched: A Netflix Engagement Report}}.
\newblock
\urldef\tempurl%
\url{https://about.netflix.com/en/news/what-we-watched-a-netflix-engagement-report}
\showURL{%
Retrieved March 19, 2024 from \tempurl}


\bibitem[Poddar(2020)]%
        {nowplaying-RS}
\bibfield{author}{\bibinfo{person}{Asmita Poddar}.} \bibinfo{year}{2020}\natexlab{}.
\newblock \bibinfo{title}{nowplaying-RS: Music Recommendation using Factorization Machines}.
\newblock
\newblock
\urldef\tempurl%
\url{https://github.com/asmitapoddar/nowplaying-RS-Music-Reco-FM}
\showURL{%
\tempurl}


\bibitem[Retailrocket(2023)]%
        {Retailrocket}
\bibfield{author}{\bibinfo{person}{Retailrocket}.} \bibinfo{year}{2023}\natexlab{}.
\newblock \bibinfo{booktitle}{\emph{Retailrocket recommender system dataset}}.
\newblock Retailrocket.
\newblock
\urldef\tempurl%
\url{https://www.kaggle.com/datasets/retailrocket/ecommerce-dataset}
\showURL{%
\tempurl}


\bibitem[Rice(1976)]%
        {RICE197665}
\bibfield{author}{\bibinfo{person}{John~R. Rice}.} \bibinfo{year}{1976}\natexlab{}.
\newblock \showarticletitle{The Algorithm Selection Problem**This work was partially supported by the National Science Foundation through Grant GP-32940X. This chapter was presented as the George E. Forsythe Memorial Lecture at the Computer Science Conference, February 19, 1975, Washington, D. C.}
\newblock \bibinfo{series}{Advances in Computers}, Vol.~\bibinfo{volume}{15}. \bibinfo{publisher}{Elsevier}, \bibinfo{pages}{65--118}.
\newblock
\showISSN{0065-2458}
\urldef\tempurl%
\url{https://doi.org/10.1016/S0065-2458(08)60520-3}
\showDOI{\tempurl}


\bibitem[Romero et~al\mbox{.}(2013)]%
        {romero2013meta}
\bibfield{author}{\bibinfo{person}{Cristobal Romero}, \bibinfo{person}{Juan~Luis Olmo}, {and} \bibinfo{person}{Sebasti{\'a}n Ventura}.} \bibinfo{year}{2013}\natexlab{}.
\newblock \showarticletitle{A meta-learning approach for recommending a subset of white-box classification algorithms for Moodle datasets}. In \bibinfo{booktitle}{\emph{Educational Data Mining 2013}}.
\newblock


\bibitem[Rosenthal(2017)]%
        {Sketchfab}
\bibfield{author}{\bibinfo{person}{Ethan Rosenthal}.} \bibinfo{year}{2017}\natexlab{}.
\newblock \bibinfo{title}{Sketchfab: content discovery... IN 3D}.
\newblock
\newblock
\urldef\tempurl%
\url{https://github.com/EthanRosenthal/rec-a-sketch}
\showURL{%
\tempurl}


\bibitem[spotifiy.com({[n.\,d.]})]%
        {spotify_report}
\bibfield{author}{\bibinfo{person}{spotifiy.com}.} \bibinfo{year}{[n.\,d.]}\natexlab{}.
\newblock \bibinfo{booktitle}{\emph{About Spotify}}.
\newblock
\urldef\tempurl%
\url{https://newsroom.spotify.com/company-info/}
\showURL{%
Retrieved March 19, 2024 from \tempurl}


\bibitem[Sun et~al\mbox{.}(2019)]%
        {10.1145/3357384.3357895}
\bibfield{author}{\bibinfo{person}{Fei Sun}, \bibinfo{person}{Jun Liu}, \bibinfo{person}{Jian Wu}, \bibinfo{person}{Changhua Pei}, \bibinfo{person}{Xiao Lin}, \bibinfo{person}{Wenwu Ou}, {and} \bibinfo{person}{Peng Jiang}.} \bibinfo{year}{2019}\natexlab{}.
\newblock \showarticletitle{BERT4Rec: Sequential Recommendation with Bidirectional Encoder Representations from Transformer}. In \bibinfo{booktitle}{\emph{Proceedings of the 28th ACM International Conference on Information and Knowledge Management}} (Beijing, China) \emph{(\bibinfo{series}{CIKM '19})}. \bibinfo{publisher}{Association for Computing Machinery}, \bibinfo{address}{New York, NY, USA}, \bibinfo{pages}{1441–1450}.
\newblock
\showISBNx{9781450369763}
\urldef\tempurl%
\url{https://doi.org/10.1145/3357384.3357895}
\showDOI{\tempurl}


\bibitem[Vente(2023)]%
        {10.1145/3604915.3608886}
\bibfield{author}{\bibinfo{person}{Tobias Vente}.} \bibinfo{year}{2023}\natexlab{}.
\newblock \showarticletitle{Advancing Automation of Design Decisions in Recommender System Pipelines}. In \bibinfo{booktitle}{\emph{Proceedings of the 17th ACM Conference on Recommender Systems}} (Singapore, Singapore) \emph{(\bibinfo{series}{RecSys '23})}. \bibinfo{publisher}{Association for Computing Machinery}, \bibinfo{address}{New York, NY, USA}, \bibinfo{pages}{1355–1360}.
\newblock
\showISBNx{9798400702419}
\urldef\tempurl%
\url{https://doi.org/10.1145/3604915.3608886}
\showDOI{\tempurl}


\bibitem[Vente et~al\mbox{.}(2023)]%
        {10.1145/3604915.3610656}
\bibfield{author}{\bibinfo{person}{Tobias Vente}, \bibinfo{person}{Michael Ekstrand}, {and} \bibinfo{person}{Joeran Beel}.} \bibinfo{year}{2023}\natexlab{}.
\newblock \showarticletitle{Introducing LensKit-Auto, an Experimental Automated Recommender System (AutoRecSys) Toolkit}. In \bibinfo{booktitle}{\emph{Proceedings of the 17th ACM Conference on Recommender Systems}} (Singapore, Singapore) \emph{(\bibinfo{series}{RecSys '23})}. \bibinfo{publisher}{Association for Computing Machinery}, \bibinfo{address}{New York, NY, USA}, \bibinfo{pages}{1212–1216}.
\newblock
\showISBNx{9798400702419}
\urldef\tempurl%
\url{https://doi.org/10.1145/3604915.3610656}
\showDOI{\tempurl}


\bibitem[Wang et~al\mbox{.}(2022)]%
        {10.1145/3485447.3512147}
\bibfield{author}{\bibinfo{person}{Jianling Wang}, \bibinfo{person}{Ya Le}, \bibinfo{person}{Bo Chang}, \bibinfo{person}{Yuyan Wang}, \bibinfo{person}{Ed~H. Chi}, {and} \bibinfo{person}{Minmin Chen}.} \bibinfo{year}{2022}\natexlab{}.
\newblock \showarticletitle{Learning to Augment for Casual User Recommendation}. In \bibinfo{booktitle}{\emph{Proceedings of the ACM Web Conference 2022}} (<conf-loc>, <city>Virtual Event, Lyon</city>, <country>France</country>, </conf-loc>) \emph{(\bibinfo{series}{WWW '22})}. \bibinfo{publisher}{Association for Computing Machinery}, \bibinfo{address}{New York, NY, USA}, \bibinfo{pages}{2183–2194}.
\newblock
\showISBNx{9781450390965}
\urldef\tempurl%
\url{https://doi.org/10.1145/3485447.3512147}
\showDOI{\tempurl}


\bibitem[Wang et~al\mbox{.}(2020)]%
        {10.1145/3383313.3411529}
\bibfield{author}{\bibinfo{person}{Ting-Hsiang Wang}, \bibinfo{person}{Xia Hu}, \bibinfo{person}{Haifeng Jin}, \bibinfo{person}{Qingquan Song}, \bibinfo{person}{Xiaotian Han}, {and} \bibinfo{person}{Zirui Liu}.} \bibinfo{year}{2020}\natexlab{}.
\newblock \showarticletitle{AutoRec: An Automated Recommender System}. In \bibinfo{booktitle}{\emph{Proceedings of the 14th ACM Conference on Recommender Systems}} (Virtual Event, Brazil) \emph{(\bibinfo{series}{RecSys '20})}. \bibinfo{publisher}{Association for Computing Machinery}, \bibinfo{address}{New York, NY, USA}, \bibinfo{pages}{582–584}.
\newblock
\showISBNx{9781450375832}
\urldef\tempurl%
\url{https://doi.org/10.1145/3383313.3411529}
\showDOI{\tempurl}


\bibitem[Xu and Tian(2015)]%
        {xu2015comprehensive}
\bibfield{author}{\bibinfo{person}{Dongkuan Xu} {and} \bibinfo{person}{Yingjie Tian}.} \bibinfo{year}{2015}\natexlab{}.
\newblock \showarticletitle{A comprehensive survey of clustering algorithms}.
\newblock \bibinfo{journal}{\emph{Annals of Data Science}}  \bibinfo{volume}{2} (\bibinfo{year}{2015}), \bibinfo{pages}{165--193}.
\newblock
\urldef\tempurl%
\url{https://doi.org/10.1007/s40745-015-0040-1}
\showDOI{\tempurl}


\bibitem[Xue et~al\mbox{.}(2005)]%
        {10.1145/1076034.1076056}
\bibfield{author}{\bibinfo{person}{Gui-Rong Xue}, \bibinfo{person}{Chenxi Lin}, \bibinfo{person}{Qiang Yang}, \bibinfo{person}{WenSi Xi}, \bibinfo{person}{Hua-Jun Zeng}, \bibinfo{person}{Yong Yu}, {and} \bibinfo{person}{Zheng Chen}.} \bibinfo{year}{2005}\natexlab{}.
\newblock \showarticletitle{Scalable collaborative filtering using cluster-based smoothing}. In \bibinfo{booktitle}{\emph{Proceedings of the 28th Annual International ACM SIGIR Conference on Research and Development in Information Retrieval}} (Salvador, Brazil) \emph{(\bibinfo{series}{SIGIR '05})}. \bibinfo{publisher}{Association for Computing Machinery}, \bibinfo{address}{New York, NY, USA}, \bibinfo{pages}{114–121}.
\newblock
\showISBNx{1595930345}
\urldef\tempurl%
\url{https://doi.org/10.1145/1076034.1076056}
\showDOI{\tempurl}


\bibitem[Yue et~al\mbox{.}(2021)]%
        {10.1145/3460231.3474275}
\bibfield{author}{\bibinfo{person}{Zhenrui Yue}, \bibinfo{person}{Zhankui He}, \bibinfo{person}{Huimin Zeng}, {and} \bibinfo{person}{Julian McAuley}.} \bibinfo{year}{2021}\natexlab{}.
\newblock \showarticletitle{Black-Box Attacks on Sequential Recommenders via Data-Free Model Extraction}. In \bibinfo{booktitle}{\emph{Proceedings of the 15th ACM Conference on Recommender Systems}} (Amsterdam, Netherlands) \emph{(\bibinfo{series}{RecSys '21})}. \bibinfo{publisher}{Association for Computing Machinery}, \bibinfo{address}{New York, NY, USA}, \bibinfo{pages}{44–54}.
\newblock
\showISBNx{9781450384582}
\urldef\tempurl%
\url{https://doi.org/10.1145/3460231.3474275}
\showDOI{\tempurl}


\bibitem[Yue et~al\mbox{.}(2022)]%
        {10.1145/3523227.3546770}
\bibfield{author}{\bibinfo{person}{Zhenrui Yue}, \bibinfo{person}{Huimin Zeng}, \bibinfo{person}{Ziyi Kou}, \bibinfo{person}{Lanyu Shang}, {and} \bibinfo{person}{Dong Wang}.} \bibinfo{year}{2022}\natexlab{}.
\newblock \showarticletitle{Defending Substitution-Based Profile Pollution Attacks on Sequential Recommenders}. In \bibinfo{booktitle}{\emph{Proceedings of the 16th ACM Conference on Recommender Systems}} (<conf-loc>, <city>Seattle</city>, <state>WA</state>, <country>USA</country>, </conf-loc>) \emph{(\bibinfo{series}{RecSys '22})}. \bibinfo{publisher}{Association for Computing Machinery}, \bibinfo{address}{New York, NY, USA}, \bibinfo{pages}{59–70}.
\newblock
\showISBNx{9781450392785}
\urldef\tempurl%
\url{https://doi.org/10.1145/3523227.3546770}
\showDOI{\tempurl}


\bibitem[zalando.com({[n.\,d.]})]%
        {zalando_report}
\bibfield{author}{\bibinfo{person}{zalando.com}.} \bibinfo{year}{[n.\,d.]}\natexlab{}.
\newblock \bibinfo{booktitle}{\emph{A holistic strategy for a connected fashion world}}.
\newblock
\urldef\tempurl%
\url{https://corporate.zalando.com/en/about-us/holistic-strategy-connected-fashion-world}
\showURL{%
Retrieved March 19, 2024 from \tempurl}


\end{thebibliography}

\end{document}